\documentclass[a4paper,11pt]{article}
\pdfoutput=1 

\usepackage{jcappub} 

\usepackage[normalem]{ulem}
\usepackage[utf8]{inputenc}
\usepackage{enumerate}
\usepackage{amsmath, amssymb, amsthm, graphicx, epsfig, fancyhdr,epsfig, slashed}
\usepackage[mathscr]{euscript}
\let\oldcdot\cdot
\usepackage{breqn}
\let\cdot\oldcdot
\usepackage{epsfig}  
\usepackage{graphicx}   
\usepackage{slashed}       
\usepackage{tikz}
\usepackage{subcaption} 
\usepackage{url}
\usepackage{color,soul} 
\usepackage{xcolor}
\usepackage{multirow} 
\usepackage{comment}
\usepackage{mathtools}
\usepackage{bm}
\usepackage{hyperref}
\usepackage[all]{hypcap}
\hypersetup{  
    colorlinks=true,
    linkcolor=blue,
    filecolor=red,      
    urlcolor=cyan,
    citecolor=red,
    }

\newcommand{\be}{\begin{eqnarray}}
\newcommand{\ee}{\end{eqnarray}}
\newcommand{\f}{\Tilde{f}}
\newcommand{\mdm}{M_{\rm DM}}
\newcommand{\mb}{M_{\rm SM}}

\begin{document}
\title{\boldmath Confronting global 21-cm signal with $\mathbb{Z}_3$ symmetric dark matter models} 

\author[a,1]{Debarun Paul,\note{Corresponding author.}}
\author[a]{Antara Dey,}
\author[a]{Amit Dutta Banik,}
\author[a,b]{Supratik Pal}


\affiliation[a]{Physics and Applied Mathematics Unit, Indian Statistical Institute, 203 B.T. Road,\\ Kolkata 700108, India}
\affiliation[b]{Technology Innovation Hub on Data Science, Big Data Analytics and Data Curation, \\ Indian Statistical Institute, 203 B.T. Road, Kolkata 700108, India}

\emailAdd{debarun31paul@gmail.com}
\emailAdd{antaraaddey@gmail.com}
\emailAdd{amitdbanik@gmail.com}
\emailAdd{supratik@isical.ac.in}

\abstract{While the $\mathbb{Z}_3$ symmetric dark matter  models have shown tremendous prospects in addressing a number of (astro-)particle physics problems, they can leave interesting imprints on cosmological observations as well. We consider two such promising models: semi-annihilating dark matter (SADM)  and Co-SIMP $2\rightarrow 3$ interaction, and investigate their effects on the global 21-cm signal. 
SADM alone cannot address the EDGES dip but can perform better with the aid of an excess radio background, whereas Co-SIMP can naturally explain the EDGES absorption feature by virtue of an intrinsic cooling effect without invoking any such excess radiation. Hence, the latter model turns out to be a rare model within the domain of CDM, that uses leptophilic interaction to achieve the EDGES dip. Further, keeping in mind the ongoing debate between EDGES and SARAS 3 on the global 21-cm signal, we demonstrate that our chosen models can still remain viable in this context, even if the EDGES data requires reassessment in future. We then extend our investigation to possible reflections on the Dark Ages, followed by a consistency check with the CMB and BAO observations via Planck 2018(+BAO) datasets.
This work thus presents a compelling case of exploring these interesting particle physics models  in the light of different cosmological observations.

}

\maketitle

\section{Introduction}
\label{sec:intro}

We are heading towards an era of cosmo-particle physics in the true sense of the term. With a couple of interesting cosmological observations already in hand and several proposed missions expected to be operational in a decade, the areas that exhibit considerable overlap of cosmology and particle physics are expected to grow tremendously. Exploring possible interplay between these two sectors is extremely essential these days, not only because of the limitations of standard cosmological models in providing fool-proof explanations to all the  observations at all scales, but also due to the limitations of ongoing particle physics experiments in providing tight constraints on the interesting particle physics processes that are being investigated nowadays. Given the scenario, the cosmological observations that are expected to come in the near future can act as a very interesting playground for these particle physics models, not only to open up possibilities to test the models from a novel perspective but also to provide possible solutions to the limitations of existing cosmological models. Either way it will be interesting to confront new cosmological experiments with new particle physics models, and vice-versa, in order to have a better understanding of both the sectors. Thus, cosmology and particle physics today should better be treated as complementary to each other. 

The baseline $\Lambda$CDM (where $\Lambda$ stands for the cosmological constant, a cosmic entity responsible for the present expansion of the Universe and CDM denotes cold dark matter) model of cosmology, despite its remarkable success in explaining a number of observations at different scales more or less satisfactorily, faces challenges in explaining certain crucial observational features such as substructure problems in small scales~\cite{Flores_1994,Oh_2011,walker2011method,2011MNRAS.415L..40B}, tensions in certain parameters like Hubble parameter $H_0$, clustering parameter $\sigma_8$~\cite{DiValentino:2021izs,Abdalla:2022yfr}, etc.
Furthermore, our understanding of the particle nature of dark matter (DM) remains elusive both from theoretical and as well as experimental perspectives, adding to the complexity of the search for a satisfactory model.
Over a period, these limitations have resulted in a growing body of works to search for  different classes of DM, e.g., warm dark matter (WDM) with DM mass $\sim$ $\mathcal{O}$(keV)~\cite{Blumenthal:1984bp,Bode:2000gq}, self interacting dark matter (SIDM)~\cite{Spergel:1999mh,Vogelsberger:2014pda,Tulin:2017ara}, decaying DM~\cite{Wang:2014ina}, fuzzy DM with DM mass $\sim$ $\mathcal{O}$($10^{-22}$eV)~\cite{Hui:2016ltb,Du:2016zcv} to name a few. However,  all of them bear their own set of limitations~\cite{Maccio:2012qf,Menci:2016eui,Irsic:2017ixq}.

The measurement of the global 21-cm signal provides valuable information about the nature of the Universe during intermediate redshifts $(100\gtrsim z \gtrsim 2)$ which is still relatively unexplored. The Experiment to Detect the Global Epoch of Reionization Signature (EDGES)~\cite{Bowman:2018yin} collaboration is such an experiment which reported a sky-averaged absorption signal around redshift $z\sim 17$. The measured differential brightness temperature, was found to be $T_{21}=-500^{+200}_{-500}$ mK~\cite{Bowman:2018yin}. However, this measurement of the extra dip in temperature showed a significant tension of 3.8$\sigma$ with the predictions of the standard $\Lambda$CDM scenario~\cite{Barkana:2018lgd}. The differential brightness temperature ($T_{21}$) which is a measure of the temperature of the 21-cm signal that originates from the hyperfine transition between the singlet ($S=0$) and triplet ($S=1$) states of ground state hydrogen atom, is mainly proportional to $\left(1-\frac{T_{\gamma}}{T_s}\right)$~\cite{FURLANETTO2006181,2012RPPh...75h6901P}. Here $T_{\gamma}$ and $T_s$ represent the radiation and neutral hydrogen spin temperatures which strongly depend on the gas kinetic temperature and Lyman-$\alpha$ temperature around $z\sim 17$, respectively. Hence, in order to reconcile the EDGES observation (an absorption signal resulting in a significant dip in the differential brightness temperature), we need to either increase the radiation temperature or decrease the spin temperature via a decrease in gas temperature. One approach is to include an excess radio background~\cite{fixsen2011arcade, 2012JAI.....150004T,Subrahmanyan:2013eqa,Dowell:2018mdb} in addition to the cosmic microwave background (CMB) radiation, a topic extensively investigated in the literature \cite{Fialkov:2019vnb,Reis:2020arr,Feng:2018rje,Sikder:2023ysk}. The second solution relies on any process by which the gas temperature can be cooled down. This has led the community to implement novel interaction between DM and visible matter~\cite{Fialkov:2018xre, Munoz:2018pzp, Munoz:2018jwq, Mukhopadhyay:2021shy,Aboubrahim:2021ohe,Barkana:2022hko}. 

The two processes above motivate us to explore a specific scenario where DM exhibits preservation of a special symmetry, namely, $\mathbb{Z}_3$ symmetry, thereby acting as a plausible source of energy-exchange between the dark and standard sectors.  While the simplest choice of DM preserving $\mathbb{Z}_2$ symmetry, has been extensively studied~\cite{silveira1985scalar,mcdonald1994gauge,burgess2001minimal}, it is known to have certain  shortcomings~\cite{cline2013update,feng2015closing}. Going beyond $\mathbb{Z}_2$, one can arrive at this interesting $\mathbb{Z}_3$ symmetry that can be preserved  in the standard sector~\cite{VanDong:2022rox,koerich2015dark} as well as in dark sector~\cite{Belanger:2012zr,Bernal:2015lbl,DiazSaez:2022nhp}. 
If a scalar singlet DM (say, $X$) is preserved under $\mathbb{Z}_3$ group, then $X$ remains invariant under $\mathbb{Z}_3$ transformation $X\rightarrow \text{exp}(i2 \pi/3)X$, which stabilizes the DM. 
The presence of $\mathbb{Z}_3$ symmetry in the DM model opens up intriguing possibilities for new and interesting processes that distinguish itself from $\mathbb{Z}_2$ symmetric DM. One such process is known as semi-annihilation~\cite{DEramo:2010keq,DEramo:2011rlb,hambye2009hidden,Belanger:2012vp,Belanger:2014bga,Bandyopadhyay:2022tsf,Ghosh:2020lma} where partial annihilation of two DM particles produces a single standard model (SM) particle along with a DM particle (SADM models henceforth). Another possible interaction allowed by $\mathbb{Z}_3$ symmetry, is Co-SIMP, a fascinating variant of SIMP (strongly interacting massive particle) dark matter~\cite{Hochberg:2014dra,Hochberg:2014kqa,Teplitz:2000zd,Mohapatra:1999gg} which has garnered attention and has been a subject of recent particle physics research~\cite{Smirnov:2020zwf,Parikh:2023qtk}. 
It is noteworthy that these $\mathbb{Z}_3$ symmetric DM models can also potentially address the small scale issues by virtue of a sizable DM self-interaction, thereby overcoming major limitations of the $\Lambda$CDM model. Nevertheless, prospects of probing these models, specially Co-SIMP processes, via different terrestrial experiments like Beam Dump experiments~\cite{Marsicano:2018vin,Batell:2014mga}, electron $g-2$ measurement~\cite{Muong-2:2021ojo}, etc, make them more interesting in recent times. 
Consequently, it draws our attention towards an explanation of these models in the light of different cosmological observations. Furthermore, their non-trivial interactions and energy-exchange with the standard sector may leave non-trivial signatures at different scales of cosmology that are worth probing. In this article, we would mostly focus on the impact of both SADM and Co-SIMP models on 21-cm cosmology using publicly available code \texttt{RECFAST}~\cite{Seager:1999bc}. As it will turn out, both the models show prospects in explaining the depth of the global 21-cm signal from the EDGES data. Although SADM alone cannot explain the EDGES absorption feature, it can do so when excess radiation (experimentally supported by ARCADE-2 and LWA1 experiments~\cite{fixsen2011arcade, 2012JAI.....150004T,Subrahmanyan:2013eqa,Dowell:2018mdb}) are taken into consideration. What makes the Co-SIMP model truly fascinating is its ability to explain the EDGES dip~\cite{Bowman:2018yin} due to the intrinsic cooling effect of the model, without the need of any excess radio background as such. This special feature of Co-SIMP model positions them as a compelling candidate for elucidating the EDGES result within the CDM framework.

However, recently a similar mission named Shaped Antenna measurement of the background RAdio Spectrum 3 (SARAS 3) has reported a rebuttal of the EDGES signal at $95.3\%$ confidence level~\cite{Singh:2021mxo, Bevins:2022ajf}, leading to an ongoing debate around the EDGES observation. The mismatch between these two promising missions may be settled either by a joint analysis of the two, including systematics, or with the help of a new dataset from the upcoming missions like Square Kilometre Array (SKA)~\cite{2015aska.confE.174B}, Lunar Surface Electromagnetics Explorer (LuSEE Night)~\cite{bale2023lusee}, Hydrogen Epoch of Reionization Array (HERA)~\cite{DeBoer:2016tnn}, Murchison Widefield Array (MWA)~\cite{Tingay:2012ps}, New Extension in Nanay Upgrading LOFAR (NenuFAR)~\cite{8471648}, Radio Experiment for the Analysis of Cosmic Hydrogen (REACH)~\cite{deLeraAcedo:2022kiu}, Mapper of the IGM Spin Temperature (MIST)~\cite{MIST}, etc. Keeping this in mind, we further demonstrate that even if EDGES results need reassessments in future, our chosen DM models can still provide a valid alternatives to $\Lambda$CDM respecting the consistency with present cosmological data. This analysis showcases the versatility of the chosen DM models, making them resilient to potential revisions of the EDGES data.

In order to search for further prospects of the models, we carry forward our analysis to other possible eras. In anticipation of the upcoming LuSEE Night~\cite{bale2023lusee} (capable of probing up to $z\sim 100$), Dark Ages Polarimeter PathfindER (DAPPER)~\cite{Chen:2019xvd} (capable of probing up to $z\sim 80$), Probing ReionizATion of the Universe using Signal from Hydrogen (PRATUSH)~\cite{pratush} observations, etc., we have examined the effects of  our chosen models on the second trough, occurring at around $z\sim 85$ of the brightness temperature. As  the astrophysical parameters have no influence during this epoch, it allows the brightness temperature to predominantly reflect the properties of the DM. By studying the behavior of the brightness temperature in the second trough we can gain insights into the characteristics and properties of the DM model. This analysis will contribute to our understanding of the nature of DM and its impact on the evolution of the Universe at relatively high redshifts.

Further, as the CMB TT  spectra arise from the temperature fluctuations which primarily originate from matter density perturbations and metric perturbations, any DM model should be in tune with CMB and BAO (baryon acoustic oscillation) data. 
In order to test our models against those datasets, we make necessary modifications to the publicly available Boltzmann equation solver \texttt{CLASS}~\cite{2011JCAP...07..034B} so as to consistently implement our perturbation equations.
Then, with the parameter space allowed by the EDGES experiment, we employ the publicly available Markov Chain Monte Carlo (MCMC) based code, \texttt{MontePython}~\cite{Brinckmann:2018cvx}, that helps in obtaining posterior distributions for the model parameters from cosmological datasets like Planck 2018, BAO etc. The entire analysis will act as a consistency check for our models against different available cosmological observations at different eras.

The development of the article is as follows.
We begin by introducing the $\mathbb{Z}_3$ symmetric DM models of our consideration in Section \ref{sec:model}. In Section \ref{sec:diff_brightness_temp} we  focus on the impact of the DM model during the Cosmic Dawn era and find out the constraints on the model parameters from the EDGES experiment. We further discuss the ongoing debate on the EDGES results and revival of $\Lambda$CDM-like behavior from our models. Moving forward, in Section \ref{sec:dark_ages} we discuss briefly the effect of our DM models on Dark Ages. Section \ref{sec:pert-eq} is dedicated to a consistency check of our models with latest CMB and BAO data, by modifying \texttt{CLASS} and subsequent implementing in \texttt{MontePython}. Finally in Section \ref{sec:conclusion} we summarize the major findings and suggest potential avenues for future research.

\section{\texorpdfstring{$\mathbb{Z}_3$}{} symmetric dark matter models}
\label{sec:model}
As discussed, $\mathbb{Z}_3$ symmetric DM models show  promise in addressing quite a few interesting particle physics problems.
$\mathbb{Z}_3$ symmetry is realized as a discrete symmetry of cyclic group of order three. Generators of this abelian $\mathbb{Z}_3$ symmetry are $\omega,~\omega^2,~1$ where $\omega^3=1$. The particular cases of our interest would be those where dark matter is charged under $\mathbb{Z}_3$ symmetry 
whereas the visible sector is neutral under $\mathbb{Z}_3$. This  imposed $\mathbb{Z}_3$ symmetry stabilizes the dark matter candidate. $\mathbb{Z}_3$ symmetry in the dark sector can also arise from spontaneously broken $U(1)_X$ gauge symmetry in the dark sector.
In such a scenario, the residual $\mathbb{Z}_3$
provides stability to the dark matter candidate.
Some interesting applications of $\mathbb{Z}_3$ symmetric dark matter have been discussed in the Introduction section.

In this article, we have handpicked two such promising models to search for their cosmological signatures, in particular in addressing the global 21-cm signal and keeping consistency with the latest Planck 2018 and BAO data. 

Let us first delve into the particular models of our interest. They  involve interaction between DM and SM particles preserving an additional $\mathbb{Z}_3$ symmetry that stabilizes DM. For our analysis, we are focusing on the following two models which maintain this symmetry:
\begin{enumerate}[i)]
    \item DM semi-annihilation: $DM\,+DM \,\rightarrow\,DM\,+\,SM$  ~\cite{DEramo:2010keq,DEramo:2011rlb,hambye2009hidden,Belanger:2012vp,Belanger:2014bga,Bandyopadhyay:2022tsf,Ghosh:2020lma}.
    \item Co-SIMP $2\,\rightarrow\,3$ interaction: $DM\,+SM\rightarrow\,DM\,+DM +\,SM\,$ ~\cite{Smirnov:2020zwf,Parikh:2023qtk}. 
\end{enumerate}

These models can show potentially interesting behavior by virtue of the interaction between the dark and standard sectors. As a result, it is intriguing to investigate if they can leave significant imprints on the era of the Dark Ages and the Cosmic Dawn. This draws our attention to these models and motivates us to inspect them in the light of other cosmological observations to gain a deeper understanding of their consequences and potential observational signatures.
\subsection{Semi-annihilating dark matter (SADM)}
\label{subsec:DMmodelSADM}
\begin{figure}[ht!]
    \centering
    \includegraphics[scale=0.6]{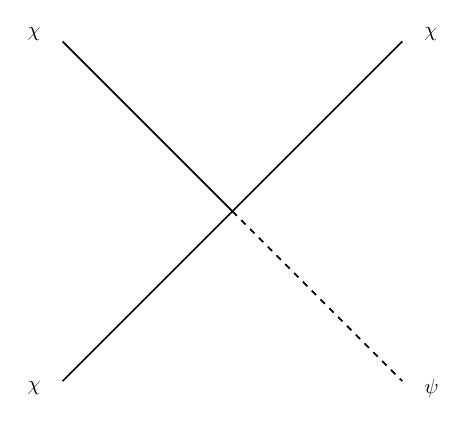}
    \caption{Diagram for semi-annihilation process. $\chi$ represents Dark Matter and $\psi$ represents Standard Model particles.}
    \label{fig:feynman_SADM}
\end{figure}

In this framework we consider a single component DM species $\chi$ preserving an additional $\mathbb{Z}_3$ symmetry which can have the following interaction with the SM species $\psi$~\cite{DEramo:2010keq,DEramo:2011rlb,hambye2009hidden,Belanger:2012vp,Belanger:2014bga,Bandyopadhyay:2022tsf,Ghosh:2020lma}, $$\chi\,+\,\chi\,\rightarrow\,\chi\,+\,\psi,$$ as depicted in Fig.~\ref{fig:feynman_SADM}.
Unlike any $2\rightarrow 2$ DM annihilation, $2\rightarrow 2$ DM semi-annihilation changes the SM species by unity, leading to a non-trivial effect on the DM relic density and heat exchange between the dark and standard sectors.
\par
The evolution of the number density of DM for this semi-annihilation process can be described via the Boltzmann equation as 
\be\label{eq:BEQ}
\frac{dY_{\rm DM} (x)}{dx} = -\frac{x\langle \sigma\,v\rangle s}{H(M_{\rm DM})}\left(Y_{\rm DM}(x)^2 - Y_{\rm DM}(x) \,Y_{eq}(x)\right),
\ee
where the co-moving number density $Y_{\rm DM}$ can be expressed as the ratio between the DM number density $n_{\rm DM}$ and the entropy density $s$, where $x$ represents the ratio between the DM mass $M_{\rm DM}$ and temperature of the Universe $T$. Further, in the Boltzmann equation above, the thermal averaged cross-section, denoted by $\langle \sigma v\rangle$, needs to be $\sim\,10^{-26} \rm{cm}^3/\rm{s}$ to satisfy the DM relic abundance~\cite{Planck:2018vyg} if this class of DM spans the whole set of DM of the Universe. 

\subsection{Co-SIMP dark matter with leptophilic interaction}
\label{subsec:DMmodel2to3}
\begin{figure}[ht!]
    \centering
    \includegraphics[scale=0.23]{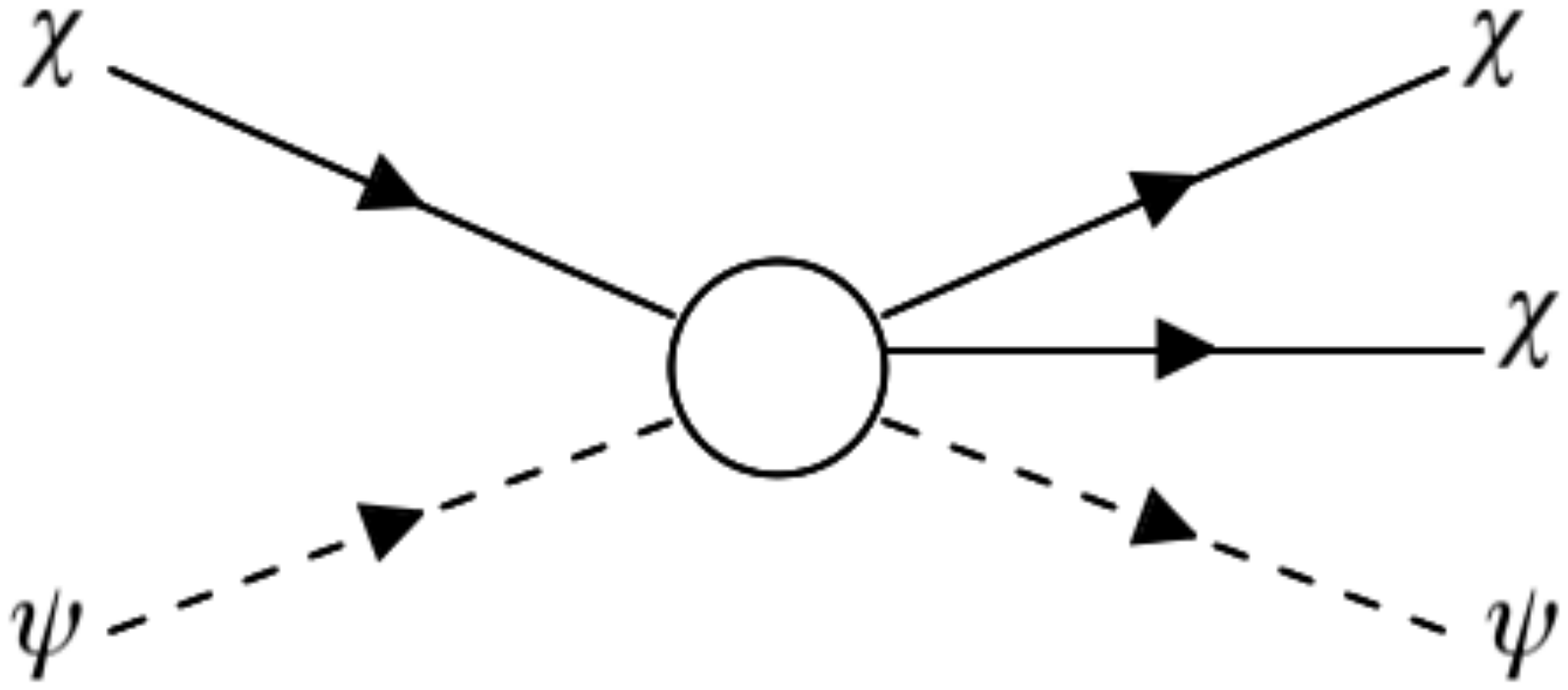}
    \caption{Schematic diagram for $2\rightarrow 3$ Co-SIMP interaction process where $\chi$ and $\psi$ represent the dark and SM particles respectively.}
    \label{fig:feynman_2to3}
\end{figure}

The second scenario that we are interested in can be depicted as DM $\chi$, which undergoes a Co-SIMP interaction~\cite{Smirnov:2020zwf,Parikh:2023qtk} with an SM species $\psi$. The specific interaction represented by a $2\rightarrow 3$ Co-SIMP  process  
$$\chi + \psi \rightarrow \chi + \chi + \psi,$$
can be represented by the following diagram in Fig.~\ref{fig:feynman_2to3}.

Here the Co-SIMP interaction is characterized by the preservation of the $\mathbb{Z}_3$ symmetry. In this article, we have focused on a leptogenic Co-SIMP interaction \cite{Parikh:2023qtk}, where the SM species $\psi$ is taken to be electron. Unlike standard scattering processes, this Co-SIMP interaction leads to an increase in the number of DM particles by unity, which has significant implications for the DM relic density and the exchange of heat between the dark and standard sectors.

\section{Differential brightness temperature of global 21-cm signal}
\label{sec:diff_brightness_temp}
The major observable in any global 21-cm experiment is the differential brightness temperature $T_{21}$, which can be expressed as
\be\label{eq:dTb}
T_{21} \simeq 27 x_{\rm HI}\left(\frac{\Omega_bh^2}{0.023}\right)\left(\frac{0.15}{\Omega_m h^2}\frac{1+z}{10}\right)^{\frac{1}{2}}\left(1-\frac{T_{\gamma}}{T_s}\right)\,\,\rm{mK},
\ee
where $x_{\rm HI}$, $T_{\gamma}$ and $T_s$ represent the neutral gas fraction, background radiation temperature and neutral hydrogen spin temperature respectively~\cite{FURLANETTO2006181,2012RPPh...75h6901P}. $T_{\gamma}$ simply equals the CMB temperature in the absence of any excess background radiation over CMB radiation and the spin temperature $T_s$ can be expressed as the weighted average of the different temperatures as follows~\cite{1958PIRE...46..240F}
\be\label{Ts}
T_s^{-1} = \frac{T_{\gamma}^{-1} + x_k T_k^{-1} + x_{\alpha}T_{\alpha}^{-1}}{1 + x_k + x_{\alpha}},
\ee
where $T_k$ and $T_{\alpha}$ represent gas kinetic temperature of the intergalactic medium and color temperature respectively.
Now $T_{\alpha} \approx T_k$ is a typical assumption as there are a high number of Lyman-$\alpha$ scattering photons that the Lyman-$\alpha$ temperature and gas temperature are brought in to thermal equilibrium~\cite{2012RPPh...75h6901P,1959ApJ...129..551F}. Hence, during the calculation of $T_s$, according to Eq.~\eqref{Ts}, we have assumed $T_{\alpha} = T_k$. Here $x_k$ and $x_{\alpha}$ represent the collisional coupling co-efficient and the Lyman-$\alpha$ coupling co-efficient respectively which have been calculated in the conventional manner as described in Refs.~\cite{Kuhlen_2006,PhysRevD.74.103502,Pritchard_2006,Barkana_2005}.\par
\subsection{Evolution of gas temperature}
\label{subsec:gas_evol}
The thermal evolution of the gas kinetic temperature $T_k$ of the intergalactic medium can be expressed as~\cite{PhysRevD.74.103502}
\be\label{eq:evolution_of_gas_temperature}
\frac{dT_k}{dz} = \frac{2T_k}{1+z} - \frac{2}{3H(z)(1+z)}\sum_i \frac{\epsilon_i}{k_B n_i}.
\ee
On the right hand side, the first term represents adiabatic cooling and the $\epsilon_i$ term incorporates the rate of energy injection or extraction per unit volume for the process $i$.\par
Two important contributions responsible for heating (or cooling) of the gas temperature are:
\begin{enumerate}[i)]
\item {\it Heating or cooling effect due to Compton scattering between electron and photon:} The rate of energy injection (or extraction) per unit volume, $\epsilon_{\rm comp}$, can be expressed as~\cite{Bharadwaj_2004} 
\be
\epsilon_{\rm comp} = \frac{3}{2}n k_B \frac{x_e}{1 + x_e + f_{He}}\frac{8 \sigma_T u_{\gamma}}{3m_e c}\left(T_{\gamma} - T_k\right)
\ee
with $x_e$, $f_{He}$, $\sigma_T$ and $u_{\gamma}$ denoting the free electron fraction, number of helium fraction, Thomson cross-section and the energy density of background photons respectively. In our analysis, we have considered $f_{He}=0.08$~\footnote{From big bang nucleosynthesis (BBN) we have mass of helium fraction, $Y_{He} = 0.2449 \pm 0.0040$~\cite{Aver:2015iza}. Hence, we can approximately write $Y_{He} = \frac{M_{He}}{M_{He} + M_H}\approx 24\%$ where $M$ is equal to the mass of atom ($m$) times number of the atom ($N$), with suffix bearing the usual meaning. As we know $m_{He}\approx 4m_H$, one can easily calculate $f_{He} = \frac{N_{He}}{N_{He} + N_H}\approx 8\%$.}~\cite{Chatterjee:2019jts}.

\item {\it X-ray heating:} The X-ray photons which are generally produced from galaxies and clusters can heat the intergalactic gas medium. The globally averaged energy injection density per unit time can be modeled as~\cite{PhysRevD.74.103502}
\be
\epsilon_X = 3.4\times 10^{33} f_{\rm heat} f_X \frac{\dot{\rho}_* (z)}{M_{\rm sun}\, \rm yr^{-1}\, \rm Mpc^{-3}} \rm J\,\rm s^{-1}\,\rm Mpc^{-3} 
\ee
where $f_{\rm heat}$ and $f_X$ are respectively the X-ray heating fraction and the normalization factor that incorporates the difference between local and high redshift observations. $\dot{\rho}_*$ is the star formation rate density which has been modeled according to Ref.~\cite{Barkana_2005}. This star formation rate density has a key dependence on a fraction, $f_*$, which represents the fraction of gas collapsed into a star.
\end{enumerate}

In our analysis, we have incorporated all the effects of the evolution of gas kinetic temperature via the publicly available code \texttt{RECFAST}~\cite{Seager:1999bc}. For this we have modified the code as per our requirements so as to consistently incorporate the effects of the two models of our interest.
\subsection{EDGES observation and role of the models}
\label{subsec:DMdeltaTb}
DM affects differential brightness temperature by injecting or absorbing the energy from the intergalactic gas by interacting with the visible sector. Depending on the DM model or the type of DM interaction, the amount of energy injection or absorption will be different. This largely affects the thermal evolution of the gas temperature for a particular set of astrophysical parameters. In the present analysis we have assumed a constant value of $f_* = 0.01$, which is in good agreement with radiation-hydrodynamic simulation for the high redshift galaxies~\cite{Schneider:2018xba,Wise_2014}. For $f_X$ and $f_{\rm heat}$ we have taken some typical values, listed in Table~\ref{tab:astoparameters}, which are physical according to Refs.~\cite{Chatterjee:2019jts,2003MNRAS.340..210G,PhysRevD.74.103502}. In Table~\ref{tab:astoparameters} we have listed all the considered values of the parameters with references.

\begin{table}[!ht]
\centering
\begin{tabular}{c| c |c}
    \hline
    \hline
    Parameter & Value & Ref.\\
    \hline
    $f_*$ & 0.01 & \cite{Schneider:2018xba,Wise_2014} \\
    $f_X$ & 0.2 & \cite{Chatterjee:2019jts,2003MNRAS.340..210G} \\
    $f_{\rm heat}$ & 0.2 & \cite{Chatterjee:2019jts,PhysRevD.74.103502}\\
    \hline
    \hline
\end{tabular}
\caption{Considered astrophysical parameters.}
\label{tab:astoparameters}
\end{table}
\subsubsection{SADM and heating of the gas}
\label{subsubsec:SADMeffect}
\begin{figure}[!ht]
    \centering
    \includegraphics[scale=0.48]{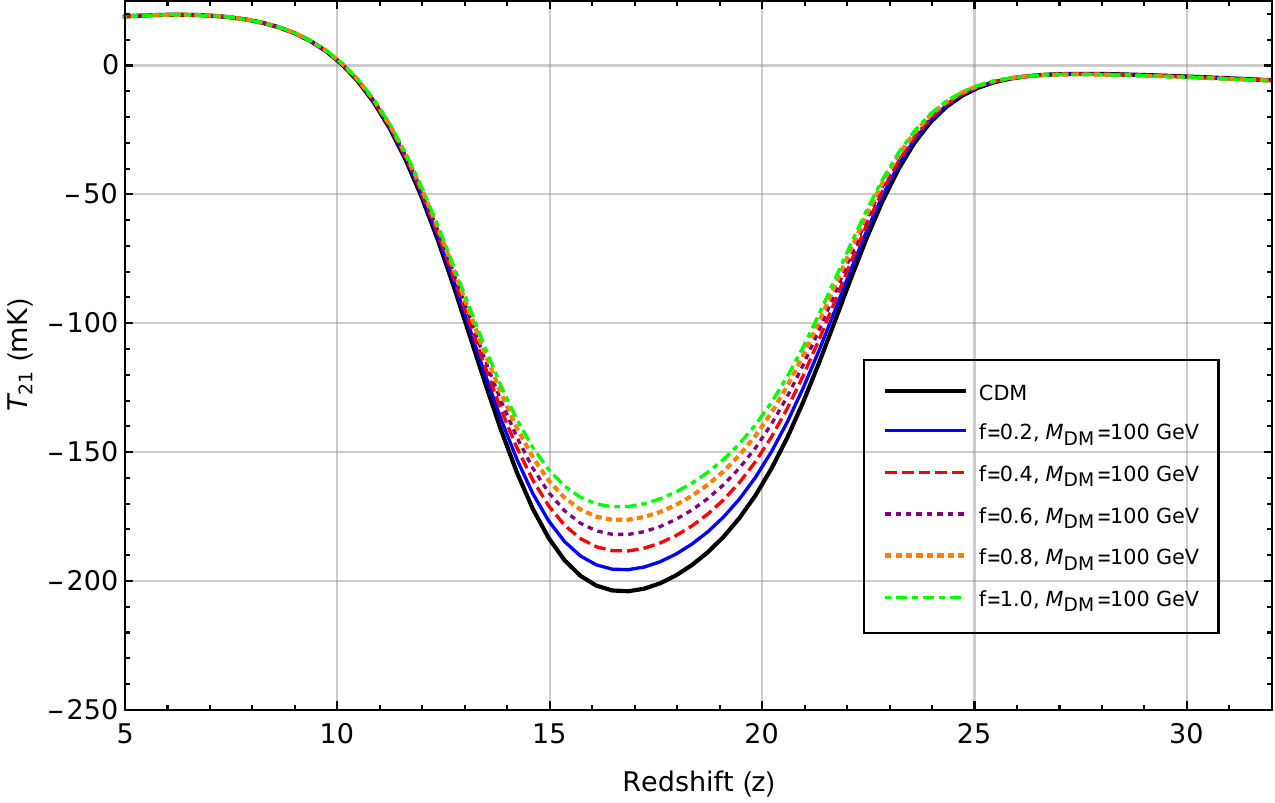}
    \caption{This figure represents the evolution of the brightness temperature ($T_{21}$) with respect to the redshift ($z$) around $z \sim 17$. It depicts that $T_{21}$ increases with the increment of $f$ and vice versa. In this analysis, we have considered $\langle \sigma v\rangle_{\rm SADM}\,=\, 3 \times 10^{-26}$ cm$^3$/s.}
    \label{fig:deltaTB_SADM_z17}
\end{figure}

Let us now discuss the possible impact of the SADM model as discussed in Section \ref{subsec:DMmodelSADM} on 21-cm cosmology. 
From Fig.~\ref{fig:feynman_SADM} it can be seen that the semi-annihilation channel of DM causes an increase in the standard sector species by one. Hence, this semi-annihilation channel injects an amount of energy equal to $f\times2M_{\rm DM}$ into the standard sector, where $f$ represents the fraction of energy transferred to the standard sector with the following constraint: $0<f<1$~\footnote{The value $f=1$ corresponds to the standard $2\rightarrow\,2$ DM annihilation scenario, in which both DM particles completely transfer their energy to the standard sector~\cite{Chluba_2010}.}. Therefore, the rate of energy injection per unit volume can be expressed as~\footnote{To calculate the energy injection rate density, we have followed the prescription given in Ref.~\cite{Chluba_2010}}
\be\label{eq:SADM_dedvdt}
\left.\frac{dE}{dVdt}\right|_{\rm SADM} = 2f \rho_{\rm DM}^2\frac{\langle \sigma v\rangle_{\rm SADM}}{M_{\rm DM}},
\ee
where $\rho_{\rm DM}$ is the density of DM, $\langle \sigma v\rangle_{\rm SADM}$ is the velocity-weighted cross section for the semi-annihilation process which is considered $3\times 10^{-26}$ cm$^3$/s, and $M_{\rm DM}$ is the mass of the DM particle. This energy injection is a function of the fractional energy transfer $f$. As a result, the evolution of differential brightness temperature $T_{21}$ and its variation with redshift $z$ will crucially depend on the value of $f$. This would be more transparent from Fig.~\ref{fig:deltaTB_SADM_z17}.\par
Fig.~\ref{fig:deltaTB_SADM_z17} illustrates the variation of brightness temperature profile with respect to redshift. Let us particularly focus on the redshift $z\sim 17$ probed by EDGES. From the figure, it is evident that $T_{21}$ increases with the increase of $f$, the fraction of heat exchange between the standard and dark sectors, thereby pushing it above $\Lambda$CDM. This in turn  makes the situation relatively worse than $\Lambda$CDM if one tries to explain the observations from EDGES~\cite{Bowman:2018yin}. This is not a big surprise as such since this is more or less generic for standard, non-interacting CDM as well as most of the annihilating DM-SM channels, as has been investigated to some extent earlier~\cite{Fialkov:2018xre, Munoz:2018pzp, Munoz:2018jwq, Mukhopadhyay:2021shy}. In fact this may be a boon  in disguise, as will be demonstrated in Section~\ref{sec:debate}.

At this moment, however, we will try to see if there is any possibility of explaining the EDGES result~\cite{Bowman:2018yin} from the SADM model. For this, we need to consider an excess background radiation on top of the CMB, which is experimentally supported by the Absolute Radiometer for Cosmology, Astrophysics and Diffuse Emission-2 (ARCADE-2) experiment~\cite{fixsen2011arcade,Subrahmanyan:2013eqa} and Long Wavelength Array (LWA1) experiment~\cite{2012JAI.....150004T,Subrahmanyan:2013eqa}. For excess radiation, we have considered the following simple model proposed in Ref.~\cite{fixsen2011arcade}:
\be\label{arcade}
T(\nu) = T_{CMB} + \xi T_R\left(\frac{\nu}{\nu_0}\right)^{\beta}
\ee
where $T_R = 1.19\pm0.14$ K, $\beta=-2.62\pm0.04$ and $\nu_0 = 1$ GHz of this model are the fitting parameters~\cite{feng2018enhanced} and $\xi$ is the free parameter which controls the amount of excess radiation in the process.\par

\begin{figure}[ht!]
    \centering
    \includegraphics[scale=0.55]{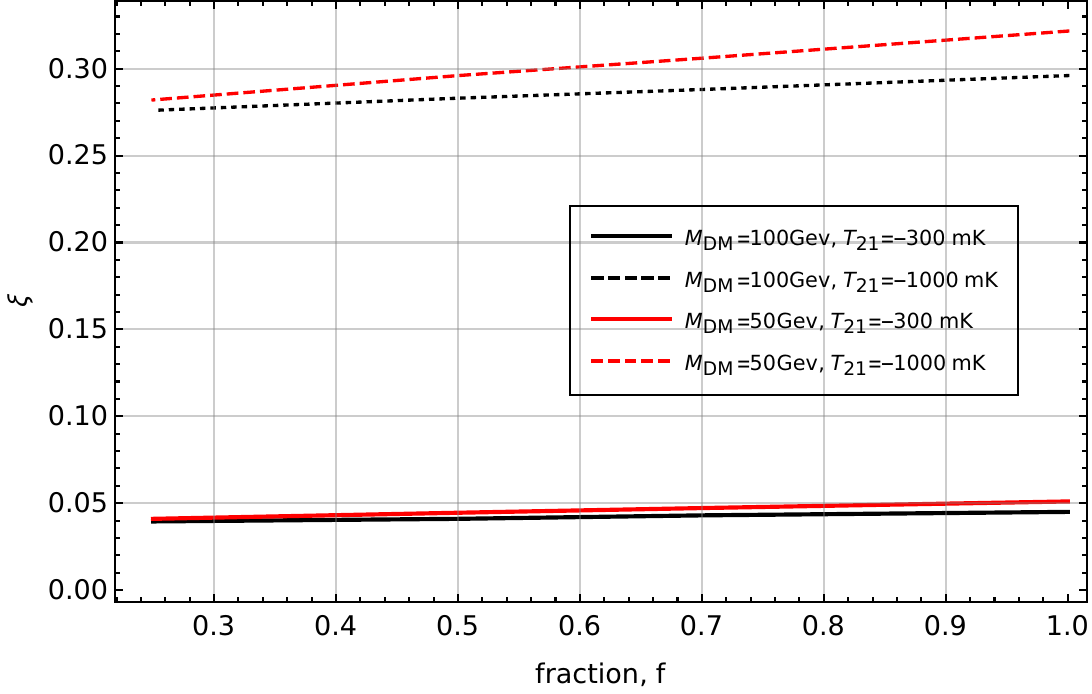}
    \caption{Variation of the amount of excess radiation, $\xi$ w.r.t. energy injection fraction ($f$). Each of the curves represents the required amount of $\xi$ to reach a desired amount of $T_{21}$ for a particular value of $f$ with a specific parameter space of self-annihilating DM}
    \label{fig:SADM4}
\end{figure}

From Eq.~\eqref{eq:SADM_dedvdt} it is obvious that for fixed values of $f$ and DM density, lower the mass of SADM greater is the amount of energy  deposited to the intergalactic medium, thereby requiring larger amount of excess radiation in order to achieve the desired level of differential brightness temperature $T_{21}$ and vice verse. This is depicted in Fig.~\ref{fig:SADM4} where we have chosen two extreme points of $T_{21}$, i.e., $-300\,{\rm mK}$ and $-1000\,{\rm mK}$, according to the EDGES observation~\cite{Bowman:2018yin}, for two specific values of DM mass 50 GeV and 100 GeV. The region in between the curves represents the allowed range of $\xi$, necessary to achieve the desired amount of $T_{21}$ for a specific DM parameter space.

At the end, we can conclude that SADM model alone appears to fail in explaining the EDGES result~\cite{Bowman:2018yin}, rather it exacerbates the situation compared to $\Lambda$CDM, as this mechanism always results in heating the gas. As in some other CDM models, this mechanism, when coupled with some excess radiation~\cite{fixsen2011arcade, 2012JAI.....150004T},  can perform better in addressing the EDGES dip. However,  this, indeed, does not reflect the credibility of the SADM mechanism itself in addressing the EDGES result, rather the role of excess radio background becomes more prudent. For that matter, it is at par with other CDM models so far as EDGES data is concerned.

\subsubsection{Co-SIMP DM-SM interaction and cooling of the gas}
\label{subsubsec:2to3effect}
In Section \ref{subsec:DMmodel2to3} we have discussed a Co-SIMP (leptogenic) interaction that increases the number of DM particles by unity. Co-SIMP processes of this kind can be probed through terrestrial searches such as Beam Dump experiments~\cite{Parikh:2023qtk} etc. However, as mentioned earlier, such an interaction could also play a significant role in 21-cm cosmology. As will be discussed in this section, this role is quite distinctive in nature. As opposed to the DM (semi)-annihilation case, this process leads to the absorption of heat from the standard sector, resulting in a decrease in the differential brightness temperature. Consequently, this unique aspect of Co-SIMP models offers an explanation for the EDGES dip~\cite{Bowman:2018yin} without the need of any excess radio background, unlike the (S)ADM case. Thus, this model makes it possible to have a viable explanation of the dip of the EDGES in terms of leptogenic interaction~\cite{Bowman:2018yin} within the CDM framework.

In our analysis, instead of taking the conventional route of incorporating the effect of DM-SM interaction for getting the evolution of gas temperature~\cite{Tashiro:2014tsa}, we have followed the energy-exchange approach between the dark and standard sectors according to Ref.~\cite{Chluba_2010}.  As DM-SM interactions exchange energy between these two species, the evolution equations for the DM and SM species can be acquired from the conservation of the total energy density of the Universe as 
\be
\Dot{\rho}_{\rm DM} + 3H\rho_{\rm DM} &=& a Q,\\
\Dot{\rho}_{\rm SM} + 3H\rho_{\rm SM} &=& -a Q,
\ee
where an overdot represents a derivative with respect to cosmic time and $Q$ represents the rate of exchange of energy density between the two species. Other terms in the expressions carry their usual meaning. As stated earlier, we have calculated the $Q$, according to the prescription given in Ref.~\cite{Chluba_2010}, expressed as
\be\label{eq:2to3_dedvdt}
Q\,\equiv\,\left.\frac{dE}{dVdt}\right|_{2\rightarrow 3} = -\f\, \sqrt{\frac{M_{\rm DM}c^2}{\left(M_{\rm SM}c^2\right)^3}}\,\sqrt{\rho_{\rm SM}^3 \rho_{\rm DM}}\,\langle \sigma v\rangle_{2\rightarrow 3},
\ee
where we have introduced $\f$, which controls the amount of energy-exchange between DM and the intergalactic gas medium. This term is responsible for decreasing or increasing the gas temperature according to Eq.~\eqref{eq:evolution_of_gas_temperature}. Here, we have considered the range of $\f$ to be $0<\f<2$ where the lower bound comes because of the presence of DM-SM interaction which is subject to exchange of energy between the two sectors and the upper bound illustrates the fact that this interaction process does not produce more than two DM particles. Similar to Eq.~\eqref{eq:evolution_of_gas_temperature}, energy transfer in Eq.~\eqref{eq:2to3_dedvdt} depends on the velocity-averaged interaction cross-section $\langle\sigma v\rangle_{2\rightarrow 3}$. As the intergalactic gas medium gets heated up by various astrophysical processes, electrons in the medium with sufficient energy can inject energy to the dark sector via the prescribed Co-SIMP mechanism. Here, $\rho$ and $M$ carry the usual meaning.  The overall negative sign in Eq.~\eqref{eq:2to3_dedvdt} depicts that the heat is absorbed by the dark sector from the standard sector. 
\par
\begin{figure}[!ht]
    \centering
    \includegraphics[scale=0.44]{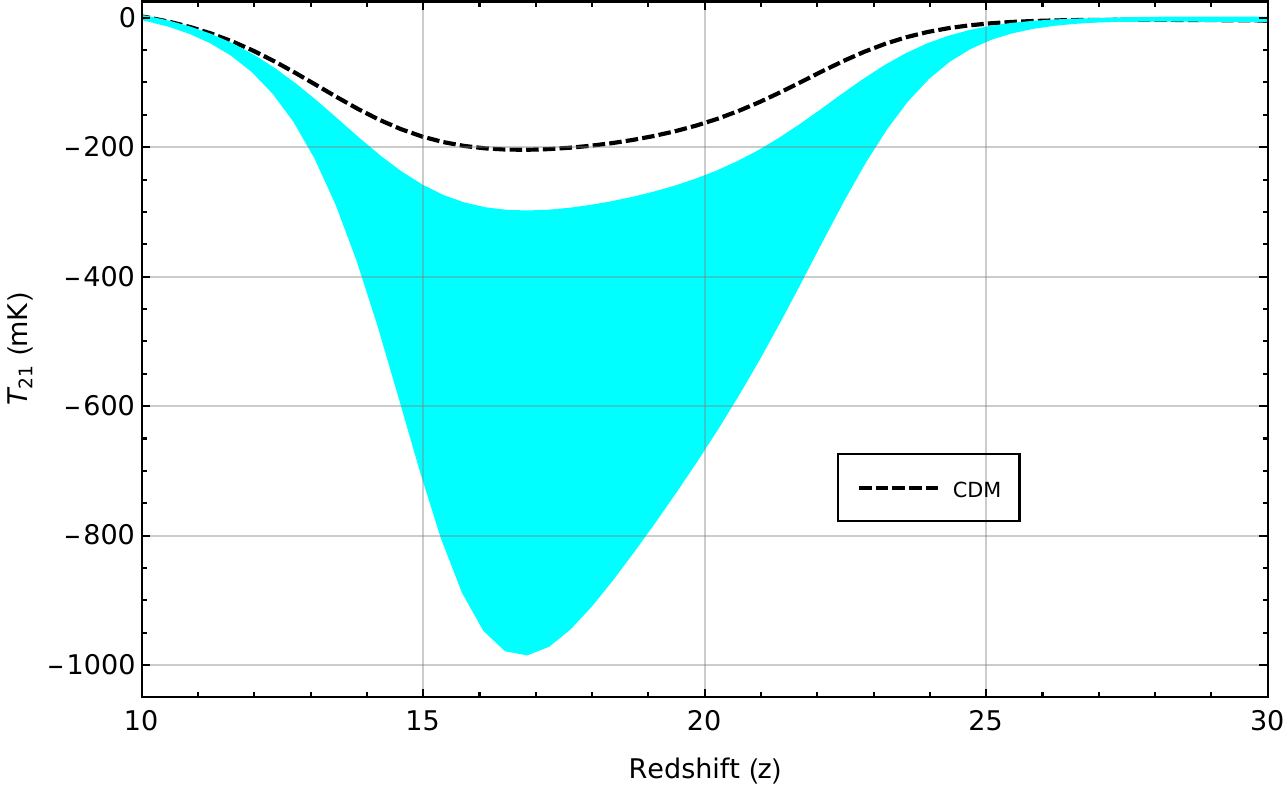}
    \caption{In this figure, the black dashed line shows the effect for the vanilla CDM model which acts a reference for the Co-SIMP process. The colored band represents the allowed region according to the observation made by the EDGES experiment~\cite{Bowman:2018yin}. This region is defined by values of $\f$ ranging from 0.56 to 1.51 and is specified by $\langle\sigma v\rangle_{2\rightarrow 3}\,=\,1.5\times 10^{-22}\, {\rm cm}^3/{\rm s}$.}
    \label{fig:delta_Tb_SADM2to3_z17}
\end{figure}
Fig.~\ref{fig:delta_Tb_SADM2to3_z17} demonstrates the variation of the differential brightness temperature for the Co-SIMP interaction compared to the case of standard, non-interacting CDM represented by the black dashed line. In our analysis we have chosen an $s$-channel interaction resulting a constant value of $\langle\sigma v\rangle_{2\rightarrow 3}$. 
To reconcile with the dip of the EDGES observation~\cite{Bowman:2018yin} we have obtained the velocity-weighted cross-section $\langle\sigma v\rangle_{2\rightarrow 3}$, which should be approximately $1.5\times 10^{-22}\,{\rm cm}^3/{\rm s}$ with $\f$ spanning from 0.56 to 1.51  shown in Fig.~\ref{fig:delta_Tb_SADM2to3_z17} in colored band, which makes it possible to successfully explain the depth of the reported absorption feature from EDGES for Co-SIMP DM model by virtue of its intrinsic cooling effect.

Let us now reiterate the most salient point of the above analysis for the Co-SIMP model. The Co-SIMP model in terms of leptonic interaction, can explain the dip in the EDGES observation~\cite{Bowman:2018yin} without the need of any excess radiation, remaining within the domain of CDM. This, in our opinion, is very noteworthy and adds a new feather to the so-called ``\textit{Co-SIMP miracle}" proposed in ~\cite{Smirnov:2020zwf}. On top of that, the rich structure of the model allows us to choose different sets of benchmark values of the model parameters and it does have further interesting consequences as will be discussed in the subsequent section. 

\subsection{Debate on EDGES  and revival of \texorpdfstring{$\Lambda$}{}CDM-like features}
\label{sec:debate}

So far we have been discussing the global 21-cm signal as measured by EDGES that cannot apparently be explained within the vanilla $\Lambda$CDM framework, and the possible role of a few interesting DM models in this context.
According to the EDGES results, the global differential brightness temperature at the Cosmic Dawn era around $z\sim 17$ is measured to be $T_{21}=-500^{+200}_{-500}$ mK~\cite{Bowman:2018yin}which indicates the presence of non-standard interactions or exotic background radiation, which is more or less twice the value predicted by the standard $\Lambda$CDM model. 

However, of late a debate is boiling on whether or not the EDGES results reveal the true global 21-cm signal at a redshift of $z\sim 17$. This is primarily because a similar mission named SARAS 3, has recently reported finding no evidence of such a signal ~\cite{Singh:2021mxo, Bevins:2022ajf} at that redshift. SARAS 3 focused on a specific redshift range $z\sim 15 - 32$ ~\cite{2021arXiv210401756N, Girish_2020} which has some overlap with EDGES low-band experiment conducted within the redshift range $z\sim 13 - 27$ ~\cite{Bowman:2018yin}. In brief, they claimed to have refuted the EDGES signal with $95.3\%$ confidence level through an independent investigation, although the exact shape of the signal still remains uncertain. 

Consequently, a substantial controversy surrounds the detection of a larger absorption signal by the EDGES experiment. There can be two possible directions to settle the issue. First, like the BICEP-2 Keck versus Planck debate on primordial gravitational waves, a possible approach to the present situation might be to engage in a joint analysis of the measurement by EDGES and SARAS teams, including possible issues with statistical analysis as well as  systematics. Secondly,  an independent measurement by a third party at the same redshift range may also help in this regard to have an independent check. 
Future experiments (e.g., SKA~\cite{2015aska.confE.174B}, LuSEE Night~\cite{bale2023lusee}, DAPPER~\cite{Chen:2019xvd}, PRATUSH~\cite{pratush} etc.) may either validate the existence of a trough with an amplitude greater than the standard value~\footnote{By `standard' we have meant the value associated to $\Lambda$CDM scenario.} of $\sim 200$ mK around $z\sim 17$~\cite{Barkana:2018lgd}, or may not find any significant dip as such. Either way it could open up exciting new perspectives into the physics at this epoch. As of now, both options are wide open.

\begin{figure*}[!ht]
    \centering
    \begin{subfigure}{.48\textwidth}
    \includegraphics[width=\textwidth]{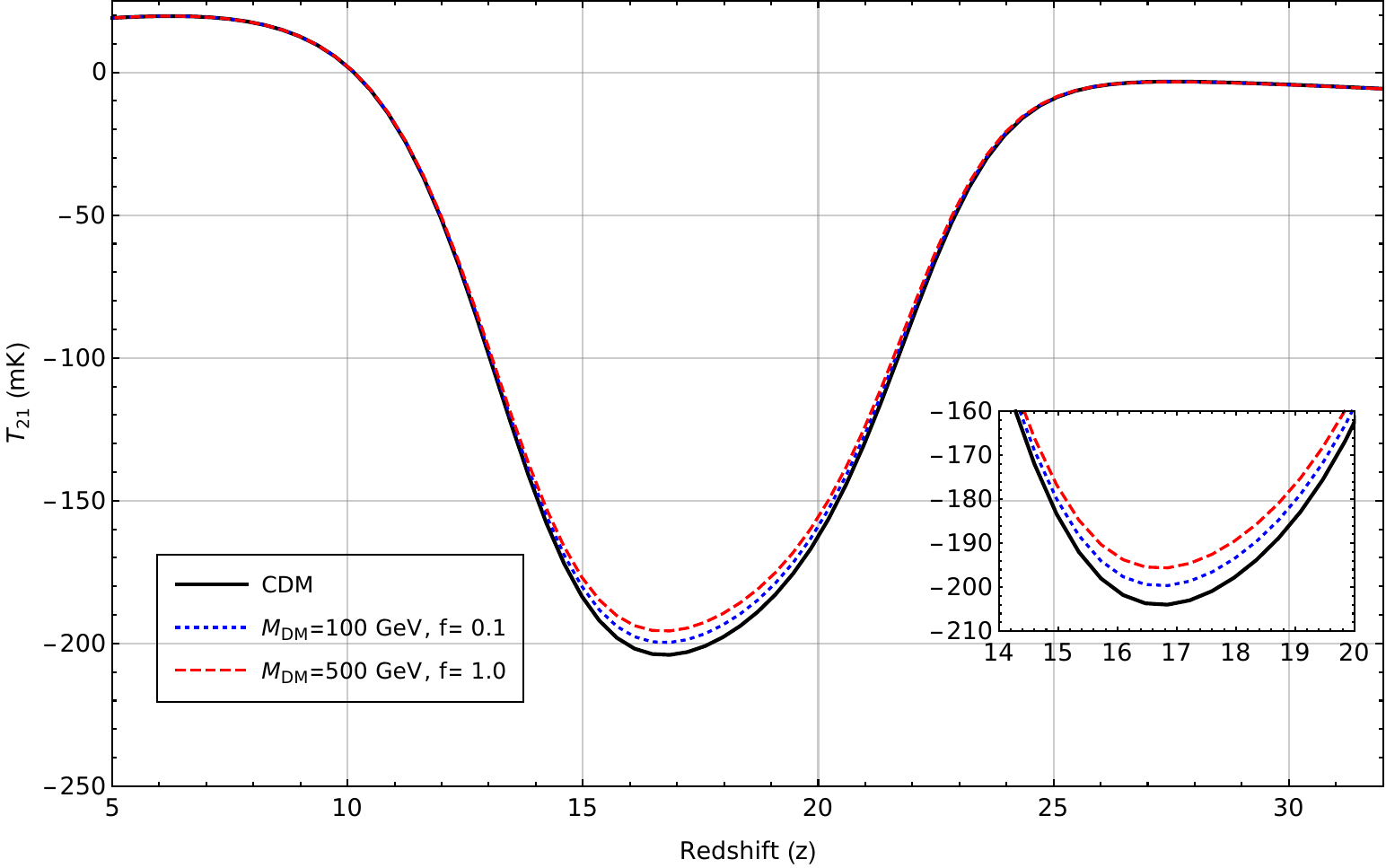}
    \caption{}
    \label{fig:SADM_vs_lcdm}
    \end{subfigure}
    \begin{subfigure}{.48\textwidth}
    \includegraphics[width=\textwidth]{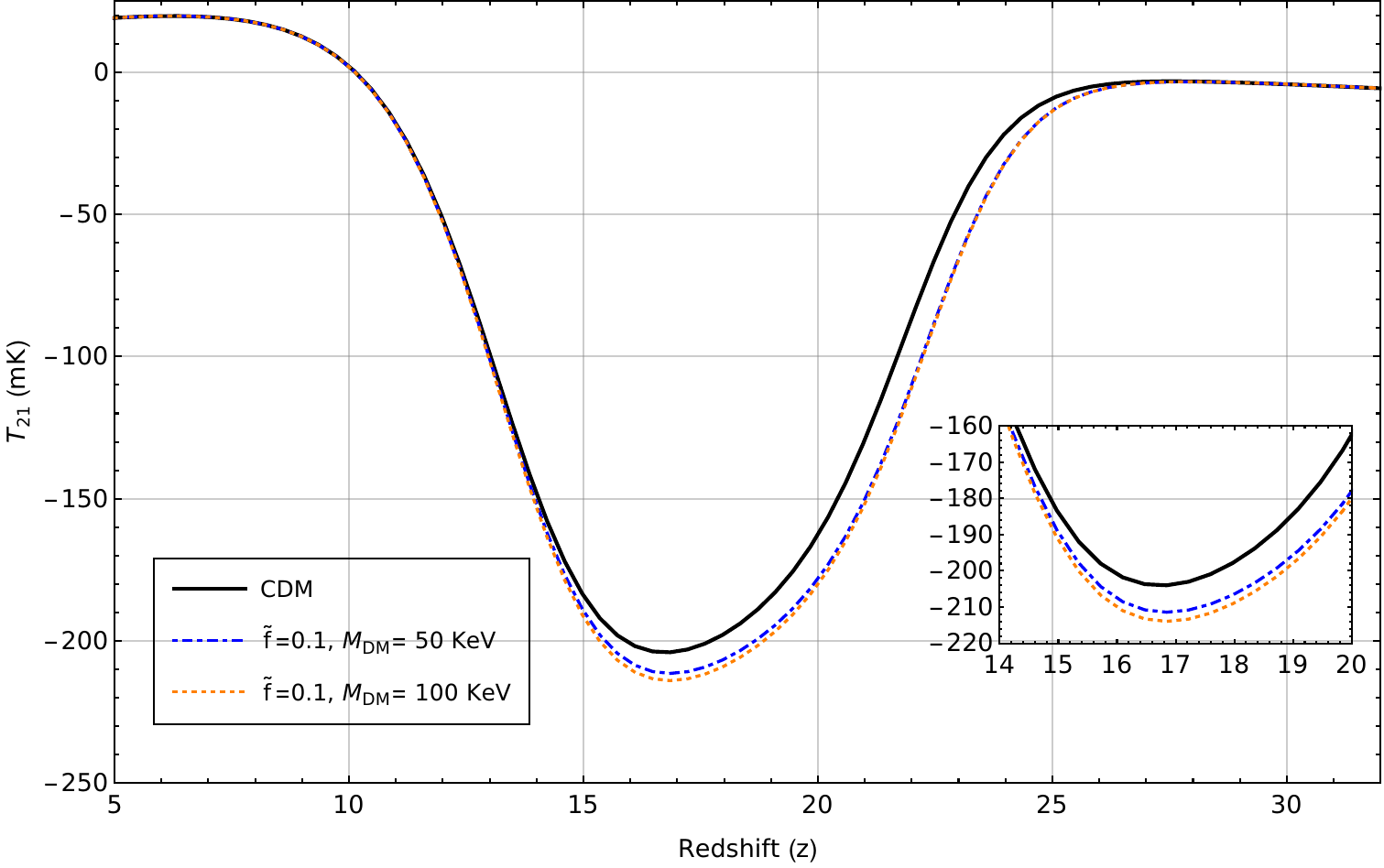}
    \caption{}
    \label{fig:cosimp_vs_lcdm}   
    \end{subfigure}
    \caption{In both of the figures, the solid black line represents the $\Lambda$CDM scenario. \textit{Left figure} illustrates the two scenarios which leads the SADM model towards vanilla $\Lambda$CDM scenario. For each set of SADM parameters we have considered $\langle\sigma v\rangle = 3.0\times 10^{-26}$ cm$^3$/s. In the \textit{Right figure} we have depicted two sets of parameters for $2\rightarrow 3$ Co-SIMP scenario which shows similar behavior as the $\Lambda$CDM scenario. For both scenarios we have considered $\langle\sigma v\rangle_{2\rightarrow 3} = 1.0\times 10^{-22}$ cm$^3$/s.}
    \label{fig:comparison_w_lcdm}
\end{figure*}

At this crucial juncture whatever model(s) one proposes in order to explain the dip, one should at the same time examine if the model can help in reviving the features close to $\Lambda$CDM. While a trivial approach to this is to switch off the effect of the model at the input level,  a wiser approach  might be to find out the sustainability of the model(s) against future data. Keeping this in mind, we have tried to see if we can get back $\Lambda$CDM-like features from the $\mathbb{Z}_3$ symmetric models of our consideration, without  switching off their effects a priori.

Although for certain values of the parameters our chosen DM models can address the dip in the EDGES observation, they can exhibit behavior close to standard $\Lambda$CDM for some other choices of model parameters as shown in Fig.~\ref{fig:comparison_w_lcdm}.  Fig.~\ref{fig:SADM_vs_lcdm} illustrates how an increment of DM mass or a decrement of cross-section drives the SADM model towards $\Lambda$CDM scenario (In Fig.~\ref{fig:SADM_vs_lcdm}, we have plotted for different benchmark values). On the other hand in Fig.~\ref{fig:cosimp_vs_lcdm}, we have shown the plots for two sets of benchmark values for which Co-SIMP shows close to $\Lambda$CDM behavior~\footnote{The considered benchmark values of the DM models are well justified. The allowed mass range of SADM is same as standard WIMP varying from within GeV to TeV mass range~\cite{DEramo:2010keq,DEramo:2011rlb,Bandyopadhyay:2022tsf}. On the other hand, the considered mass for the Co-SIMP model is also consistent with its feasible mass range~\cite{Smirnov:2020zwf,Parikh:2023qtk}.}. Of course, the exercise can be performed with other set of values as well, and the difference from $\Lambda$CDM  would be subject to the sensitivity of the instrument concerned. Given the relatively large error bar for EDGES, the major challenge any future mission has to face is to arrive at a very precise value for the differential brightness temperature at $z \sim 17$ and around. While that may not be guaranteed at this moment, that our models do have the potential to deviate from $\Lambda$CDM by a considerable amount and also show close to  $\Lambda$CDM-like behavior, due to the flexibility of the parameters concerned, makes it a strong case for them in the context of cosmology.
Hence, even if the current findings from the EDGES observation are to be reassessed or revised due to some reasons~\cite{Bevins:2022ajf,Singh:2017gtp,Patra_2013}, our chosen models would remain viable and relevant in the field of cosmology. These models offer alternative explanations and possibilities, demonstrate their resilience and continued relevance within the context of cosmological investigations.

\section{Possible reflections on Dark Ages}
\label{sec:dark_ages}
As we find interesting features of these models at Cosmic Dawn, it is intriguing to investigate if there are any other non-trivial signatures of these models at any other era. To this end, our point of interest in this section would be even higher redshifts and probe the Dark Ages~\cite{Mondal:2023xjx}. This era, which represents the epoch between last scattering of cosmic photons with baryons and formation of first stars ($1100\gtrsim z \gtrsim 30$), is still a relatively less explored era with huge prospects. As in this era the Universe is solely dominated by neutral hydrogen free of any significant astrophysical interference as such, this epoch can emerge as a marvelous probe of the nature of DM. Recognizing the significance of this epoch, such as experiments like LuSEE Night~\cite{bale2023lusee}, PRATUSH~\cite{pratush}, etc. have been proposed with the hope of probing the Universe up to $z\sim 100$. 
Hence a search for possible reflections of our models on the Dark Ages would not only lead to interesting findings but also act as an additional probe of these DM models. \par

\begin{figure*}[!ht]
    \centering
    \begin{subfigure}{.48\textwidth}
    \includegraphics[width=\textwidth]{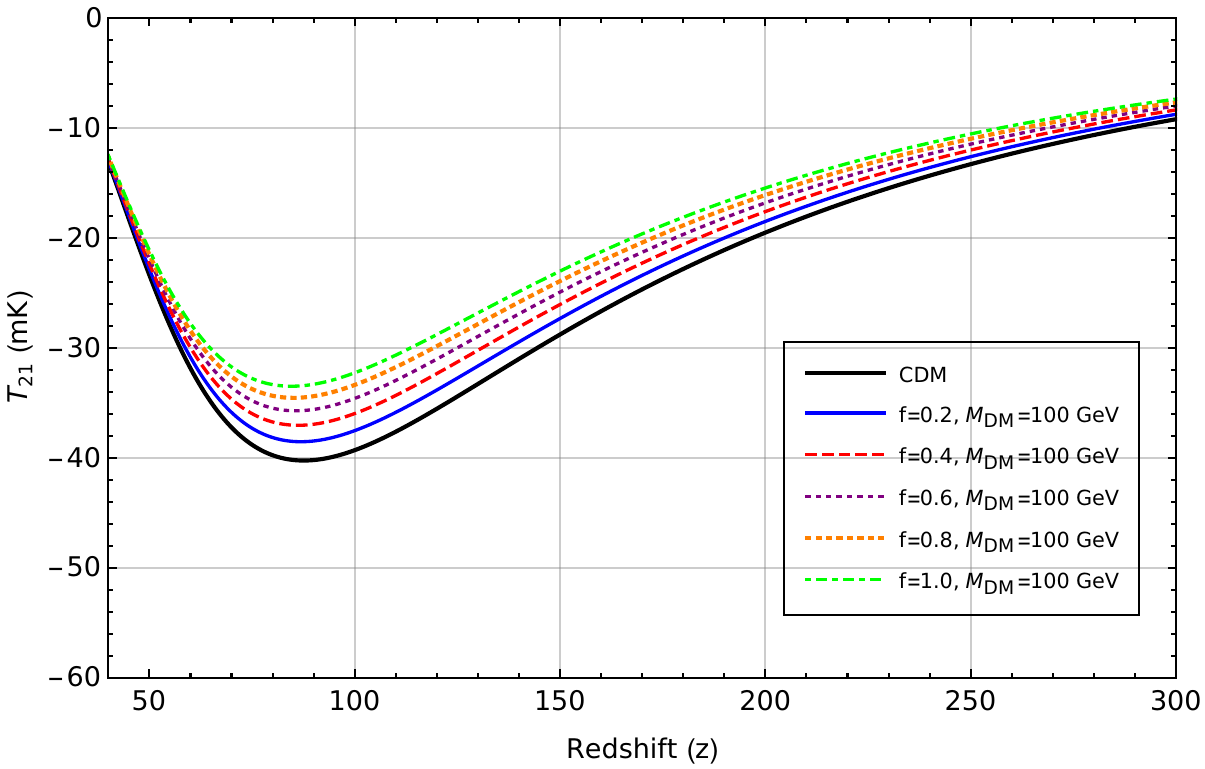}
    \caption{}
    \label{fig:deltaTB_SADM_z85}  
    \end{subfigure}
    \begin{subfigure}{.48\textwidth}
    \includegraphics[width=\textwidth]{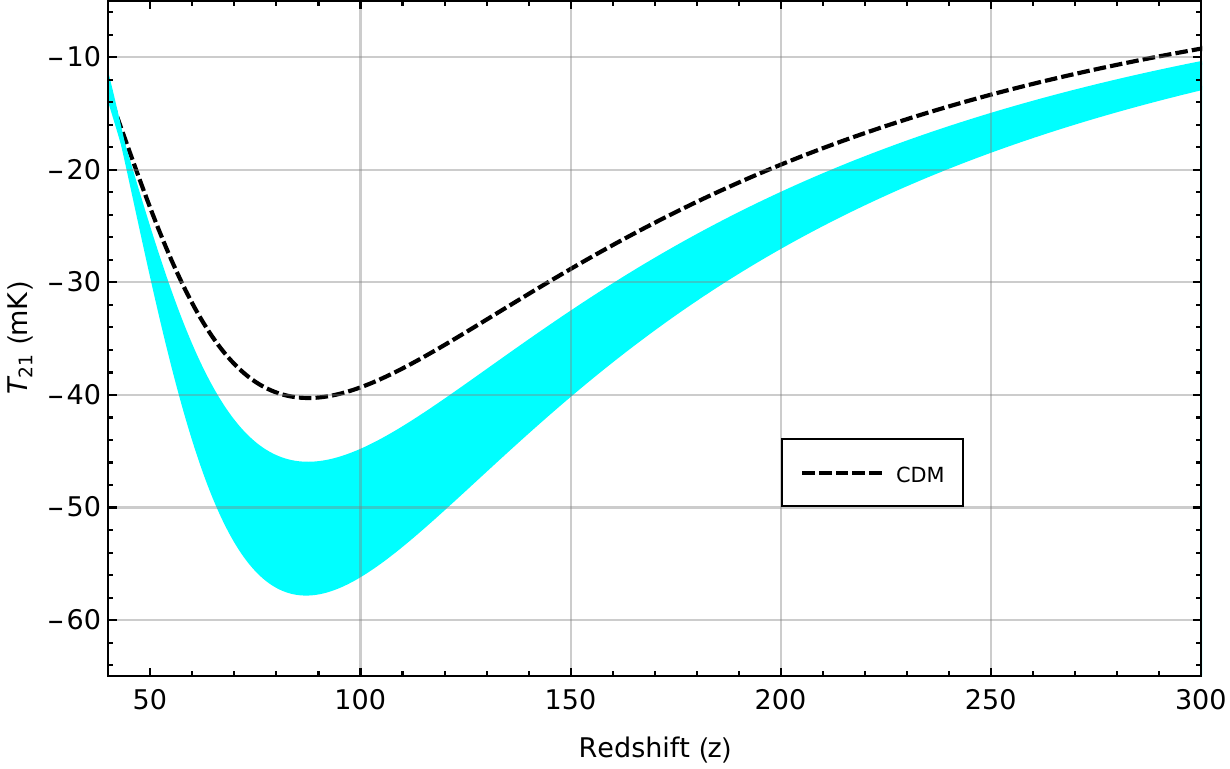}
    \caption{}
    \label{fig:delta_Tb_SADM2to3_z85}   
    \end{subfigure}
    \caption {Both figures are representative of the variation of differential brightness temperature w.r.t. redshift around redshift $z\sim 85$. \textit{Left figure} illustrates the behavior for SADM case for different values of $f$ compared with the CDM case. \textit{Right figure} depicts the same but for the Co-SIMP interaction. Here, the colored band represents the range of $\f$ allowed by the depth reported by EDGES measurement~\cite{Bowman:2018yin}, specified by $\langle\sigma v\rangle_{2\rightarrow 3}\,=\,1.5\times 10^{-22}\, {\rm cm}^3/{\rm s}$ which is compared with the CDM case depicted by black dashed line in figure.}
    \label{fig:delta_Tb_z85}
\end{figure*}

With this in mind, we have investigated the role of both of the DM models of our consideration during this era. In Fig.~\ref{fig:delta_Tb_z85} we have depicted their effects around $z\sim 85$ via the variation of differential brightness temperature with respect to redshift, keeping consistency with bounds on the model parameters as obtained from the global 21-cm signal in Sec.~\ref{subsubsec:2to3effect}.
Fig.~\ref{fig:deltaTB_SADM_z85} and Fig.~\ref{fig:delta_Tb_SADM2to3_z85} demonstrate the effects of SADM and Co-SIMP models respectively. For the case of SADM (Fig.~\ref{fig:deltaTB_SADM_z85}), we have turned off the effect of excess radio background at $z = 20$. This rise of the excess radiation can be physically explained by a rapid heating via cosmic rays~\cite{Jana:2018gqk} or a transient population of Pop-III stars~\cite{Mebane:2020jwl}. Hence, at redshift around $z\sim 85$, there will be no effect of excess radiation and the Dark Ages reflect the pure flavor of DM at that era. Fig.~\ref{fig:deltaTB_SADM_z85} reveals that the  differential brightness temperature decreases with the increase of $f$. This is  quite expected, as any increase in  $f$ induces more  absorption of energy by the gas from DM. On the other hand, the colored band in Fig.~\ref{fig:delta_Tb_SADM2to3_z85} represents the variation of $\f$ as allowed by the depth as per the EDGES observation (recall Sec.~\ref{subsubsec:2to3effect}). For this figure, it transpires that the differential brightness temperature decreases with an increase of $\f$. From a physical perspective, this can be readily explained, as an increase in $\f$ induces energy transfer from the gas to the dark species (evident from Eq.~\ref{eq:2to3_dedvdt}) for this case.\par

Both of the figures reflect that the chosen DM models have the potential to exhibit a considerable amount of deviation from the standard $\Lambda$CDM scenario but in mutually opposite directions. Nevertheless, it also manifests in the form of significant difference between the two $\mathbb{Z}_3$ symmetric models under consideration, thereby making this era potentially more interesting. This fact opens up the possibility of probing these DM models in future experiments~\cite{bale2023lusee,Chen:2019xvd,pratush} dedicated to the Dark Ages. For sure, whether or not the differences fall within the instrumental sensitivities of particular missions is subject to further investigation. However, that we have a couple of models in our hand that can successfully explain the  global 21-cm signal at $z \sim 17$ (or absence of it) and at the same time leave non-trivial imprints at Dark Ages, is indeed an interesting revelation that calls for further study with these models for the sake of next-generation cosmology and particle physics.

\section{Consistency check with other cosmological observations}
\label{sec:pert-eq}

Having demonstrated the possible role of SADM and Co-SIMP DM models on the Cosmic Dawn as well as Dark Ages, we are now in a position to check whether or not the models are consistent with the currently available cosmological datasets from other observations. This exercise has a two-fold motivation. Firstly in order for a prospective model to survive, it must pass through the experimental tests of data available from all possible eras. Secondly, it may act as an additional window to corroborate the models with observations other than 21-cm observation.  

To this end, we have first examined possible effects of those DM models on the CMB TT power spectra and matter power spectra as well as on BAO surveys. The CMB TT spectra are primarily influenced by temperature fluctuations, which arise from density perturbations thereby reflecting on the gravitational potential $\psi$ at the last scattering surface. 
These models alter the number of DM particles and also introduce a DM-SM interaction cross-section, which may reflect on density perturbations and velocity drag.  
However, since SADM essentially matches with standard non-interacting CDM at relatively higher redshifts, the perturbation equations for SADM would effectively boil down to that of CDM and one would expect that it would automatically be consistent with CMB and BAO data. 

The effects of Co-SIMP DM on perturbation equations could be rather non-trivial. So to reduce excess notation, from now we will use $\langle \sigma v\rangle$ instead of $\langle \sigma v\rangle_{2\rightarrow 3}$ to represent the velocity-averaged Co-SIMP interaction cross-section.
In Newtonian gauge, the perturbation equations for Co-SIMP DM is modified to: 
\begin{dgroup}[compact]
\begin{dmath}
\Dot{\delta}_{\rm DM} = -\theta_{\rm DM} + 3\Dot{\phi} + 2a^2\psi\sqrt{\frac{\mdm}{\mb^3}}\sqrt{\frac{\rho^{3\;(0)}_{\rm SM}}{\rho^{(0)}_{\rm DM}}}\; \langle \sigma v\rangle \;\f - \frac{1}{2}a^2\delta_{\rm DM}\sqrt{\frac{\mdm}{\mb^3}}\sqrt{\frac{\rho^{3\;(0)}_{\rm SM}}{\rho^{(0)}_{\rm DM}}}\; \langle \sigma v\rangle \;\f + a^2\sqrt{\frac{\mdm}{\mb^3}}\sqrt{\frac{\rho^{3\;(0)}_{\rm SM}}{\rho^{(0)}_{\rm DM}}}\; \delta\langle \sigma v\rangle \;\f + \frac{3}{2}a^2\sqrt{\frac{\mdm}{\mb^3}}\sqrt{\frac{\rho^{3\;(0)}_{\rm SM}}{\rho^{(0)}_{\rm DM}}}\; \delta_{\rm SM} \;\langle \sigma v\rangle \;\f\\
\end{dmath}
\begin{dmath}
\Dot{\theta}_{\rm DM} =  - H\theta_{\rm DM} + k^2\psi - a^2\theta_{\rm DM}\sqrt{\frac{\mdm}{\mb^3}}\sqrt{\frac{\rho^{3\;(0)}_{\rm SM}}{\rho^{(0)}_{\rm DM}}}\; \langle \sigma v\rangle \;\f\\
\end{dmath}
\begin{dmath}
\Dot{\delta}_{\rm SM} = -\theta_{\rm SM} + 3\Dot{\phi} - 2a^2\psi\sqrt{\frac{\mdm}{\mb^3}}\sqrt{\rho^{(0)}_{\rm SM} \rho^{(0)}_{\rm DM}}\; \langle \sigma v\rangle \;\f - \frac{1}{2} a^2 \sqrt{\frac{\mdm}{\mb^3}}\sqrt{\rho^{(0)}_{\rm SM} \rho^{(0)}_{\rm DM}}\; \langle \sigma v\rangle \;\delta_{\rm SM}\;\f - a^2\sqrt{\frac{\mdm}{\mb^3}}\sqrt{\rho^{(0)}_{\rm SM} \rho^{(0)}_{\rm DM}}\; \delta\langle \sigma v\rangle \;\f - \frac{1}{2}a^2\sqrt{\frac{\mdm}{\mb^3}}\sqrt{\rho^{(0)}_{\rm SM} \rho^{(0)}_{\rm DM}}\; \delta_{\rm DM} \;\langle \sigma v\rangle \;\f\\
\end{dmath}
\begin{dmath}
\Dot{\theta}_{\rm SM} = k^2\psi - H\theta_{\rm SM} + c_s^2k^2\delta_{\rm SM} + \frac{4\rho^{(0)}_{\gamma}}{3\rho^{(0)}_{\rm SM}}a n_e \sigma_T (\theta_{\gamma} - \theta_{\rm SM}) + a^2\theta_{\rm SM}\sqrt{\frac{\mdm}{\mb^3}}\sqrt{\rho^{(0)}_{\rm SM} \rho^{(0)}_{\rm DM}}\; \langle \sigma v\rangle \f
\end{dmath}
\end{dgroup}
where zero in the prefix represents the background (i.e. unperturbed) quantity. For simplicity to parameterize the interaction cross-section $\langle \sigma v\rangle$ in a dimensionless form, we have introduced a parameter $\Gamma_{\rm int}$, defined as
\be
\sqrt{\frac{\mdm/\mdm^{(r)}}{\mb^3/\mb^{3(r)}}}\,\frac{\langle \sigma v\rangle}{\langle \sigma v\rangle^{(r)}} \equiv \Gamma_{\rm int},
\ee
with $\mdm^{(r)}=100$ keV, $\mb^{(r)}=500$ keV and $\langle \sigma v\rangle^{(r)}=3\times 10^{-26}\,\rm{cm}^3/\rm{s}$ ~\footnote{Here by the prefix \textit{`r'}, we mean that we have set some reference value of the quantity. A constant multiplicative factor $2.39\times 10^{-25}$ is needed in front of every modifications in the code of \texttt{CLASS}~\cite{2011JCAP...07..034B} to introduce the dimensionless parameter $\Gamma_{\rm int}$. This pre-factor ensures dimensional compatibility with the \texttt{CLASS}~\cite{2011JCAP...07..034B} code.}. Hence, in this model we introduce two additional parameters, $\f$ and $\Gamma_{\rm int}$ on top of the 6 parameters of standard $\Lambda$CDM scenario. To investigate their effects on power spectra, we have implemented necessary modifications to the publicly available code \texttt{CLASS}~\cite{2011JCAP...07..034B}. Furthermore, we have utilized publicly available \texttt{MontePython} code~\cite{Brinckmann:2018cvx}, which employs the MCMC algorithm, to corroborate the model with (6+2)-parameter setup separately using the Planck 2018 (high-$l$ TT+TE+EE, low-$l$ TT, low-$l$ EE) and Planck 2018+BAO datasets.

\subsection{Likelihood analysis with Planck 2018 and BAO}
\label{subsec:likelihood}
In our cosmological analysis, we have considered a comprehensive set of parameters comprising of standard 6 cosmological parameters: $\{\omega_b,\omega_{\rm dm}, \theta_s, n_s, A_s, \tau_{\rm reio} \}$ and two additional parameters: $\{\Gamma_{\rm int}, \f \}$. Attempting to comment on the performance of the model,  we have performed MCMC analyses utilizing two datasets separately: the Planck 2018 dataset (high-$l$ TT+TE+EE, low-$l$ TT, low-$l$ EE) and the Planck 2018+BAO dataset. In Table~\ref{tab:prior}, we have displayed the priors for the (6+2) model parameters.
\begin{table}[!ht]
    \centering
    \begin{tabular}{c|c}
    \hline 
    \hline
        Parameter & Prior \\
        \hline
        $100~\omega{}_{b }$ & Flat, unbounded\\ 
        $\omega{}_{\rm dm }$ & Flat, unbounded\\ 
        $100~\theta{}_{s }$ & Flat, unbounded\\ 
        ${\rm ln}(10^{10}A_{s })$ & Flat, unbounded\\ 
        $n_{s }$ & Flat, unbounded\\ 
        $\tau{}_{reio }$ & Flat, unbounded\\ 
        $\Gamma{}_{\rm int}$ & Flat, $4.47\times10^4\rightarrow4.47\times10^{12}$\\ 
        $\f$ & Flat, $0\rightarrow2$\\ 
        \hline
        \hline
    \end{tabular}
    \caption{Priors of the (6+2) parameters used for MCMC analysis.}
    \label{tab:prior}
\end{table}

\begin{figure}
    \centering
    \includegraphics[width=\textwidth,keepaspectratio]{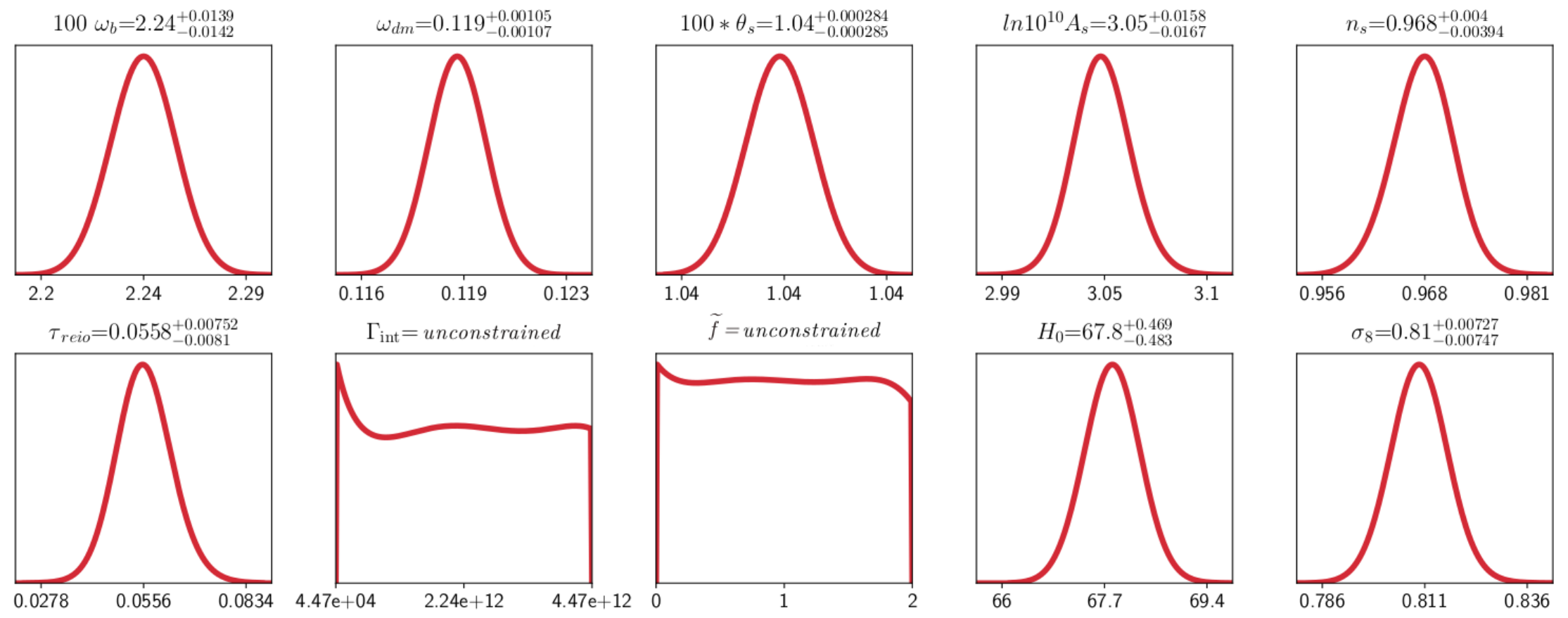}
    \caption{1-d posterior distributions for our 6+2 model $\{\omega_b,\omega_{\rm dm}, \theta_s, n_s, A_s, \tau_{\rm reio}, \Gamma_{\rm int}, \f \}$ considering Planck 2018 dataset (high-$l$ TT+TE+EE, low-$l$ TT, low-$l$ EE). The two new parameters remain unconstrained while the standard 6 parameters exhibit constraints close to $\Lambda$CDM values.}
    \label{fig:planck_1d}
\end{figure}

\begin{figure}[!ht]
    \centering
    \includegraphics[width=\textwidth,keepaspectratio]{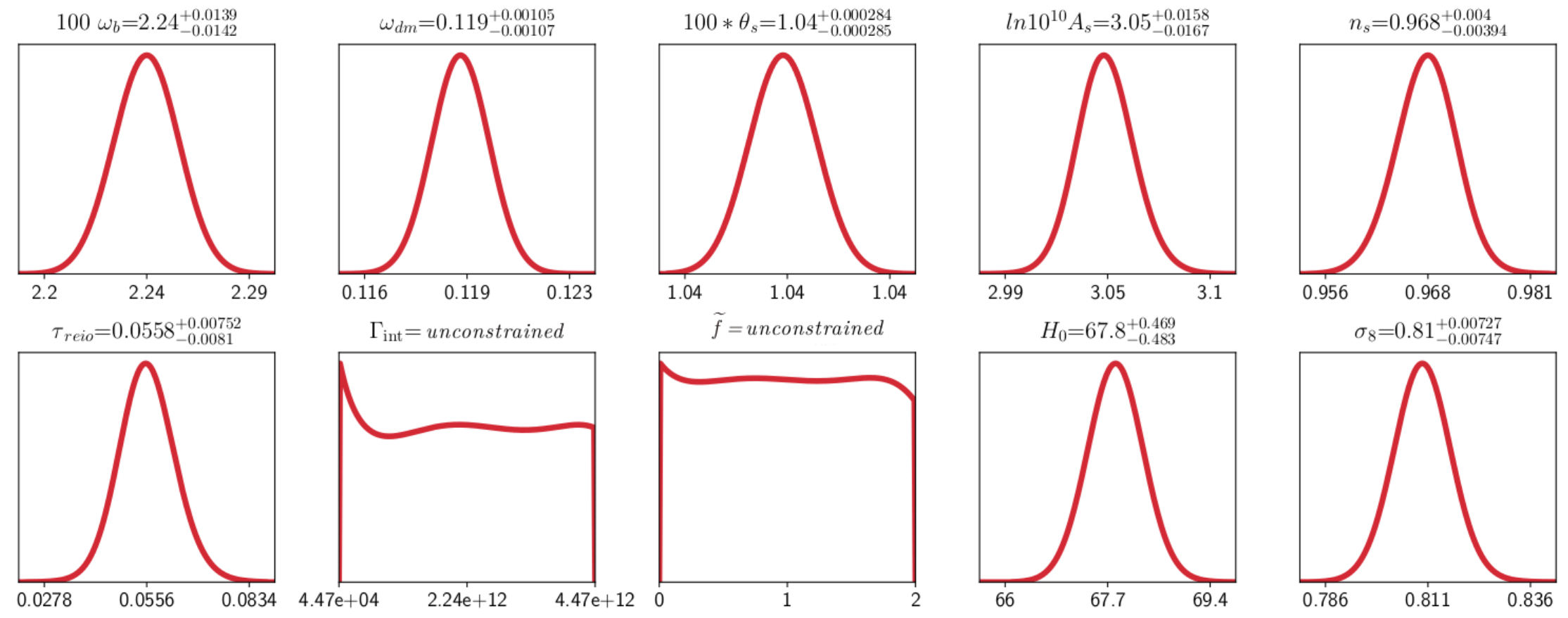}
    \caption{1-d posterior distributions for our 6+2 model $\{\omega_b,\omega_{\rm dm}, \theta_s, n_s, A_s, \tau_{\rm reio}, \Gamma_{\rm int}, \f \}$ considering Planck 2018+BAO dataset. As in Fig.~\ref{fig:planck_1d}, the two new parameters remain unconstrained while the standard 6 parameters exhibit constraints  close to $\Lambda$CDM values.}
    \label{fig:planck_bao_1d}
\end{figure}

As already mentioned, we have utilized modified versions of \texttt{CLASS}~\cite{2011JCAP...07..034B} and \texttt{MontePython} code~\cite{Brinckmann:2018cvx} to examine potential effects of the model parameters on the power spectra.
Fig.~\ref{fig:planck_1d} and Fig.~\ref{fig:planck_bao_1d} display the 1-dimensional posterior distributions for these (6+2) model parameters for Planck 2018 and Planck 2018+BAO datasets respectively. The figures reveal that the two new parameters ($\Gamma{}_{\rm int}$ and $\f$) remain mostly unconstrained by the datasets while the standard 6 parameters exhibit constraints close to $\Lambda$CDM values. 
\par

\begin{figure}[!ht]
    \centering
    \includegraphics[width=\textwidth,keepaspectratio]{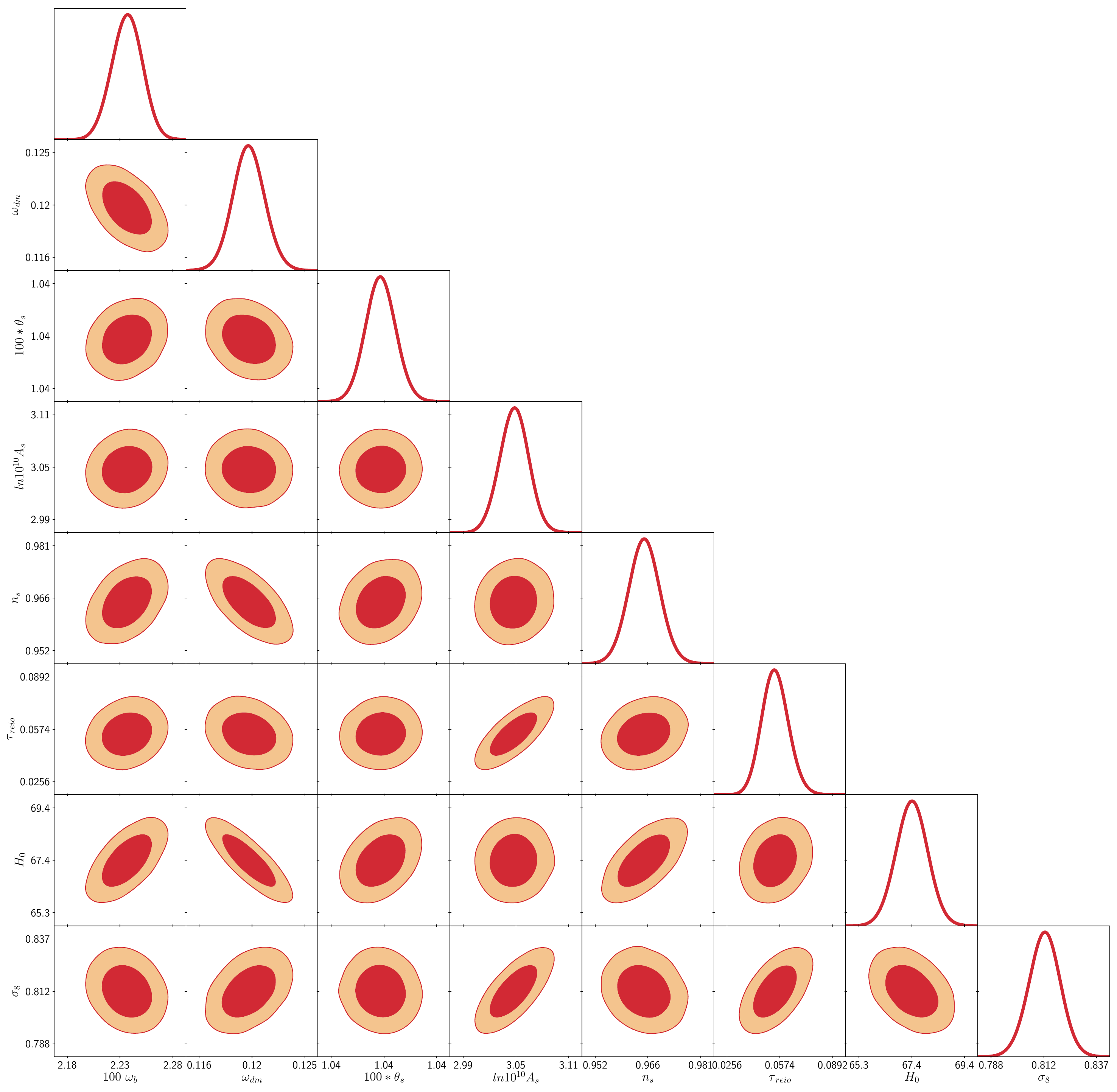}
    \caption{2-d posterior distribution for our 6 parameters $\{\omega_b,\omega_{\rm dm}, \theta_s, n_s, A_s, \tau_{\rm reio} \}$ considering Planck 2018 dataset.}
    \label{fig:planck_2d}
\end{figure}

\begin{figure}[!ht]
    \centering
    \includegraphics[width=\textwidth,keepaspectratio]{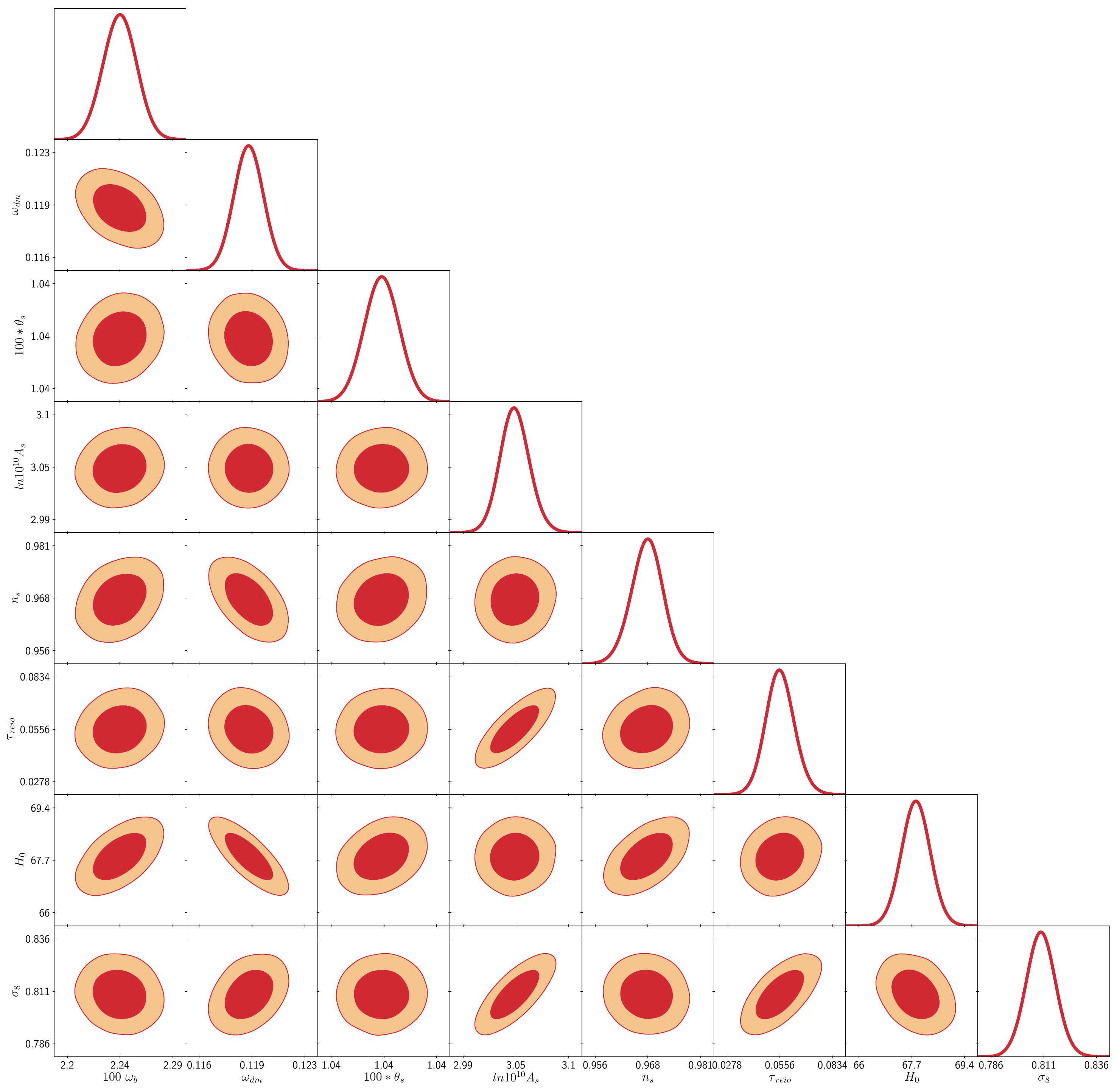}
    \caption{2-d posterior distribution for our 6 parameters $\{\omega_b,\omega_{\rm dm}, \theta_s, n_s, A_s, \tau_{\rm reio} \}$ considering Planck 2018+BAO dataset.}
    \label{fig:planck_bao_2d}
\end{figure}

As these 1-dimensional plots show that the posterior distributions for the two new parameters are nearly flat, we have shown 2-dimensional distributions only for the 6 parameters for both of the datasets in Fig.~\ref{fig:planck_2d} and Fig.~\ref{fig:planck_bao_2d}. As expected, the bounds on the standard 6 parameters are seen to be effected very little upon opening up these two parameters as one can see from Table~\ref{tab:planck_bao}, where we have shown the results of our analysis compared with the latest Planck-2018 analysis~\cite{Planck:2018vyg}.
Further, the major results of the MCMC analysis for Co-SIMP model are summarized in Table~\ref{tab:planck_bao} which illustrates the key statistical results for the posterior distributions for the considered datasets, namely the mean values of the parameters along with $68\%$ C.L.\par
\begin{table}[!ht]
    \centering
    \renewcommand{\arraystretch}{1.3}

\begin{tabular}{|p{1.7cm}|c|c|c|c|} 
 \hline 
 \hline
\multirow{3}{*}{Parameter} & \multicolumn{2}{|c|}{Planck 2018} & \multicolumn{2}{|c|}{Planck 2018+BAO} \\ \cline{2-5}
& Co-SIMP & $\Lambda$CDM & Co-SIMP & $\Lambda$CDM \\ \cline{2-5}
& mean\;$\pm \sigma$ & mean\;$\pm \sigma$ & mean\;$\pm\; \sigma$ & mean$\pm \sigma$ \\ \hline 
$100~\omega{}_{b }$ & $2.237 \pm 0.015$ & $2.236 \pm 0.015$ & $2.245 \pm 0.014$ & $2.242 \pm 0.014$\\ 
$\omega{}_{\rm dm }$ & $0.1201 \pm 0.0014$ & $0.1202 \pm 0.0014$ & $0.1191 \pm 0.0011$ & $0.11933 \pm 0.00091$\\ 
$100~\theta{}_{s }$ & $1.042_{-0.00031}^{+0.0003}$ & $1.04090 \pm 0.00031$ & $1.042 \pm 0.00028$ & $1.04101 \pm 0.00029$\\ 
${\rm ln}(10^{10}A_{s })$ & $3.046_{-0.017}^{+0.016}$ & $3.045 \pm 0.016$ & $3.046_{-0.017}^{+0.016}$ & $3.047 \pm 0.014$\\ 
$n_{s }$ & $0.9654_{-0.0046}^{+0.0044}$ & $0.9649 \pm 0.0044$ & $0.9681_{-0.0039}^{+0.004}$ & $0.9665 \pm 0.0038$\\ 
$\tau{}_{reio }$ & $0.0547_{-0.0083}^{+0.0076}$ & $0.05578_{-0.0081}^{+0.0070}$ & $0.05578_{-0.0081}^{+0.0075}$ &  $0.0561 \pm 0.0071$\\ 
$\Gamma{}_{\rm int}$ & $-$ & $-$ & $-$ & $-$\\ 
$\f$ & $-$ & $-$ & $-$ & $-$\\ 
\hline 
$H_0$ & $67.35_{-0.63}^{+0.61}$ &  $67.27 \pm 0.60$ & $67.81_{-0.48}^{+0.47}$ &  $67.66 \pm 0.42$\\ 
$\sigma_8$ & $0.8122_{-0.0078}^{+0.0074}$ &  $0.8120 \pm 0.0073$ & $0.8096_{-0.0075}^{+0.0073}$ & $0.8111 \pm 0.0060$\\

\hline 
\hline
 \end{tabular} \\
    \caption{Statistical results for our 6+2 model $\{\omega_b,\omega_{\rm dm}, \theta_s, n_s, A_s, \tau_{\rm reio}, \Gamma_{\rm int}, \f \}$ considering Planck 2018 (high-$l$ TT+TE+EE, low-$l$ TT, low-$l$ EE) and Planck 2018+BAO datasets. Except these parameters, we have also shown the statistical values of $H_0$ and $\sigma_8$ in this table. Additionally we have shown a comparison between Co-SIMP results and latest Planck-2018 results~\cite{Planck:2018vyg}, which points out that Co-SIMP model is well consistent with the latest Planck-2018 results~\cite{Planck:2018vyg}, indicating that Co-SIMP model retains the success of $\Lambda$CDM at large scale. }

    \label{tab:planck_bao}
\end{table}

That the two additional parameters for our model cannot be constrained either by Planck 2018 or by Planck 2018+BAO dataset is not a big surprise as such.
The Co-SIMP model successfully explains the dip in EDGES observation~\cite{Bowman:2018yin} by virtue of interaction between DM and energetic electrons, which is significantly dominant at the intermediate redshifts due to astrophysical effects (as discussed in section~\ref{subsubsec:2to3effect}). At the CMB scale however, due to the absence of astrophysical effects, a much smaller amount of energetic electrons will be available compared to intermediate redshifts. This leads to a nominal Co-SIMP process which is not expected to leave much trace during this epoch, leading to a flat posteriors for the additional parameters that characterize the Co-SIMP process, and almost the same constraints on the rest of the parameters as in $\Lambda$CDM. 
On the other hand, the BAO dataset is mainly sensitive to the total matter content of the Universe and the matter density contrast. In our analysis, the DM-SM interaction  does not alter any of them significantly.
Hence, the new parameters should not have significant dependence on the BAO dataset, and the values of the other cosmological parameters, including $H_0$ and $\sigma_8$, are not expected to deviate significantly from the $\Lambda$CDM scenario. 

One needs to keep in mind that, in this article, our target was not to put constraints on the two additional parameters from the model, but to check if the values of the model parameters, as obtained from the global 21-cm signal, lead to any inconsistency whatsoever at the CMB and BAO scales against the available data. That we indeed have constraints on the other cosmological parameters as expected from our understanding of cosmology, together with the fact that no additional tension appears for the two new model parameters, allows us to conclude that the model with interesting features at the reionization era is quite consistent with other cosmological data from other epochs. If, however, one wants to explore the possibility of constraining any of these two additional parameters from CMB and BAO, an interesting option  may be to include the effective number of neutrino species $N_{\rm eff}$ on top of the (6+2) parameters in this scenario. It has been shown to  provide additional constraints on low mass Co-SIMP DM which can, however, be circumvented by additional sterile neutrino species consistent with CMB and BBN \cite{Smirnov:2020zwf}. 

Further, in order to demonstrate the prospects of   $N_{\rm eff}$ in the present scenario as predicted above, we have  extended the previous MCMC analysis for the Co-SIMP model by including $N_{\rm eff}$ as an additional free parameter, thereby making it a (6+3)-parameter analysis that includes 6 cosmological parameters $\{\omega_b,\omega_{\rm dm}, \theta_s, n_s, A_s, \tau_{\rm reio} \}$ and 3 additional parameters: $\{\Gamma_{\rm int}, \f, N_{\rm eff} \}$ for Planck 2018 (high-$l$ TT+TE+EE, low-$l$ TT, low-$l$ EE) dataset. We have chosen a flat prior for $N_{\rm eff}$ ranging from $2.0$ to $5.5$. Priors for the other parameters are as same as mentioned in Table~\ref{tab:prior}. 
\begin{figure}[!ht]
    \centering
    \includegraphics[width=\textwidth,keepaspectratio]{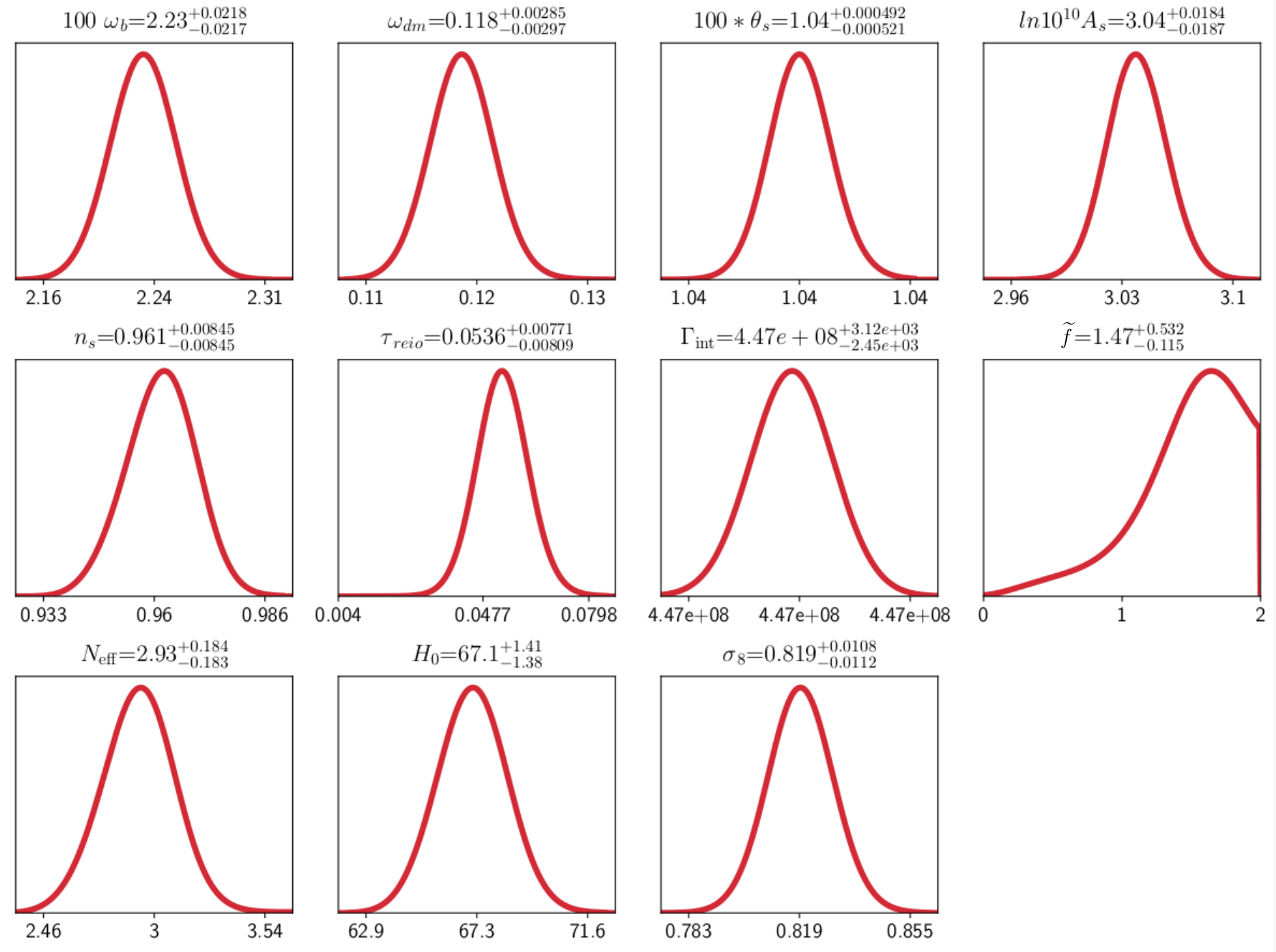}
    \caption{1-d posterior distributions for our 6+3 model $\{\omega_b,\omega_{\rm dm}, \theta_s, n_s, A_s, \tau_{\rm reio}, \Gamma_{\rm int}, \f, N_{\rm eff} \}$ considering Planck 2018 dataset (high-$l$ TT+TE+EE, low-$l$ TT, low-$l$ EE). Additionally we have included the distribution of $H_0$ and $\sigma_8$.}
    \label{fig:neff_1d}
\end{figure}

\begin{figure}[!ht]
    \centering
    \includegraphics[width=\textwidth,keepaspectratio]{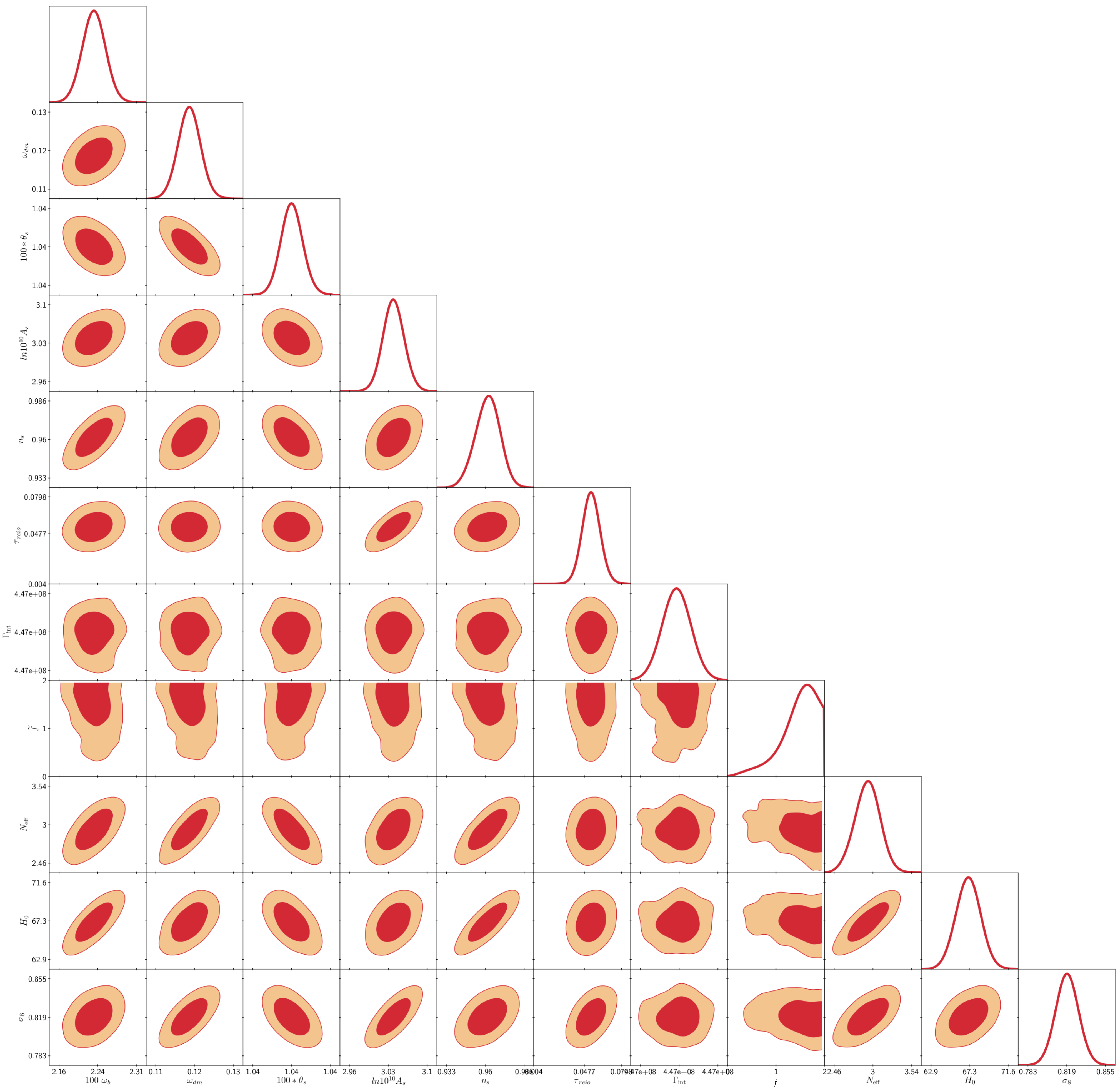}
    \caption{2-d posterior distributions for our 6+3 model $\{\omega_b,\omega_{\rm dm}, \theta_s, n_s, A_s, \tau_{\rm reio}, \Gamma_{\rm int}, \f, N_{\rm eff} \}$ using Planck 2018 dataset (high-$l$ TT+TE+EE, low-$l$ TT, low-$l$ EE). 
    }
    \label{fig:neff_2d}
\end{figure}

In Table~\ref{tab:neff_result} we have summarized the statistical results of the analysis. 1-d (Fig.~\ref{fig:neff_1d}) and 2-d posterior distribution (Fig.~\ref{fig:neff_2d})  justify that opening up effective number of neutrino species on top the (6+2) parameters in MCMC runs, indeed helps us in constraining the Co-SIMP model parameters to some extent along with producing consistent results for $N_{\rm eff}$ and other parameters, as predicted.

\begin{table}[!ht]
    \centering
    \renewcommand{\arraystretch}{1.4}
    \begin{tabular}{|l|c|c|} 
 \hline 
 \hline
Parameter & Best-fit value & mean$\pm\sigma$ \\ \hline 
$100~\omega{}_{b }$ &$2.202$ & $2.228\pm 0.022$ \\ 
$\omega{}_{\rm dm }$ &$0.1144$ & $0.1183_{-0.003}^{+0.0029}$ \\ 
$100*\theta{}_{s }$ &$1.043$ & $1.042_{-0.00053}^{+0.00049}$ \\ 
$\ln 10^{10}A_{s }$ &$3.025$ & $3.039_{-0.019}^{+0.018}$ \\ 
$n_{s }$ &$0.9561$ & $0.961\pm 0.0085$ \\ 
$\tau{}_{\rm reio}$ &$0.05308$ & $0.05356_{-0.0081}^{+0.0077}$  \\ 
$\Gamma_{\mathrm{int}}$ &$4.47e+08$ & $4.47e+08_{-2.4e+03}^{+3.1e+03}$ \\ 
$\f$ &$1.88$ & $1.48_{-0.1}^{+0.52}$ \\ 
$N_{\rm eff}$ &$2.683$ & $2.926_{-0.18}^{+0.19}$ \\
\hline
$H_{0 }$ &$65.61$ & $67.09\pm 1.4$ \\ 
$\sigma_8$ &$0.8082$ & $0.8193\pm 0.011$ \\ 
\hline 
\hline
 \end{tabular}
    \caption{Statistical result for the 6+3 model $\{\omega_b,\omega_{\rm dm}, \theta_s, n_s, A_s, \tau_{\rm reio}, \Gamma_{\rm int}, \f, N_{\rm eff} \}$ considering Planck 2018 dataset (high-$l$ TT+TE+EE, low-$l$ TT, low-$l$ EE). Except these parameters we have also displayed the statistical values of $H_0$ and $\sigma_8$ in this table.}
    \label{tab:neff_result}
\end{table}

In a nutshell, the above analysis reassures that the model is consistent with the latest Planck-2018 analysis~\cite{Planck:2018vyg}, as well as the physical processes under consideration, and at the same time can exhibit interesting features during the reionization epoch.
Moreover, this analysis is important for a consistency check with the different cosmological datasets and it is found that our chosen model is well consistent with those observations.

\section{Summary and future directions}
\label{sec:conclusion}
This article discusses the prospects of $\mathbb{Z}_3$ symmetric  DM models in the light of recent cosmological data. 
The models under consideration namely the SADM and $2\rightarrow 3$ Co-SIMP interaction between DM and SM, have been receiving a growing attention in recent times in the context of particle physics but their cosmological implications are yet unexplored. 
We have analyzed the effects of these DM models during the era known as Cosmic Dawn, with a specific emphasis on addressing the observed first trough (around $z \sim 17$) of the differential brightness temperature from the EDGES experiment~\cite{Bowman:2018yin} using the publicly available code \texttt{RECFAST}~\cite{Seager:1999bc}. 
We have shown that the SADM model cannot explain the observed absorption feature. Although with the aid of an excess of radio background~\cite{fixsen2011arcade, 2012JAI.....150004T, Dowell:2018mdb} the observed dip can be achieved using SADM model, it does not reflect credibility of the SADM mechanism. 
In contrast, the Co-SIMP interaction model shows promise in explaining the depth reported by the EDGES observation by virtue of its cooling properties without the need of any additional radio background. This feature makes Co-SIMP models stand out, offering a promising explanation in terms of leptonic interaction with DM for the EDGES observation while operating within the framework of CDM. 
Further, in order to address the  ongoing debate~\cite{Bevins:2022ajf,Singh:2017gtp,Patra_2013} with the global 21-cm signal observed by EDGES~\cite{Bowman:2018yin}, we have shown that our chosen model will still remain viable even if the EDGES results need to be reassessed. 

Furthermore, we have conducted a consistency check to ensure the compatibility of our models  with other cosmological observations. 
The publicly available code \texttt{CLASS}~\cite{2011JCAP...07..034B} (with necessary modifications so as to fit our perturbation equations) and MCMC code \texttt{Montepython}~\cite{Brinckmann:2018cvx} have been utilized to corroborate the  DM models with the latest CMB and BAO observations, separately for Planck 2018 alone and Planck 2018+BAO datasets. This allows for a comprehensive examination of the parameter space for the concerned DM models and their compatibility with observational data. MCMC results illustrate that our chosen models do not significantly affect the CMB and BAO scales,  with the standard 6 parameters getting close to $\Lambda$CDM values leading to no contradiction among different observations for these two models. We have also performed the MCMC analysis by opening up effective number of neutrino as an additional free parameter, employing solely the Planck 2018 dataset and demonstrated the prospects of $N_{\rm eff}$ in constraining the model parameters.
A further analysis on the intermediate redshifts, e.g., during the Dark Ages, reveals that these models exhibit distinctive characteristics, setting them apart not only from the standard $\Lambda$CDM scenario but also from each other. These distinctive features can serve as crucial indicators to distinguish between different DM models with the aid of future observations.

The findings of this study have implications for future experimental investigations. Upcoming Dark Ages experiments like LuSEE Night~\cite{bale2023lusee}, PRATUSH~\cite{pratush}, etc., designed to probe the Universe at redshifts around 100, can serve as an exciting platform to further constrain and scrutinize the proposed DM models. A brief analysis towards this direction has been reported in the present article. However, a rigorous investigation of these models focusing on Dark Ages is yet to be done.
Also the upcoming CMB experiments (e.g. LiteBIRD~\cite{Matsumura:2013aja,Suzuki:2018cuy}, CMB-S4~\cite{CMB-S4:2016ple,CMB-S4:2017uhf}, etc.), Large Scale Structure (LSS) experiments (e.g. Euclid~\cite{EuclidTheoryWorkingGroup:2012gxx}, Dark Energy Spectroscopic Instrument(DESI)~\cite{DESI:2013agm,DESI:2016fyo}, etc.) have tremendous potential of exploring the DM models and provide more constraints on the model parameters, allowing for a more comprehensive examination of the DM models with $\mathbb{Z}_3$ symmetry in future. We hope to explore some of them in near future.

\acknowledgments
The authors gratefully acknowledge the use of publicly available codes \texttt{RECFAST}~\cite{Seager:1999bc}, \texttt{CLASS}~\cite{2011JCAP...07..034B}, \texttt{MontePython}~\cite{Brinckmann:2018cvx} and thank the computational facilities of the Indian Statistical Institute, Kolkata. DP and AD thank ISI Kolkata for financial support through Senior Research Fellowship. ADB acknowledges financial support from DST, India, under grant number IFA20-PH250 (INSPIRE Faculty Award). SP thanks the Department of Science and Technology, Govt. of India for partial support through Grant No. NMICPS/006/MD/2020-21.

\bibliographystyle{bibi}
\bibliography{mybib.bib}

\providecommand{\href}[2]{#2}\begingroup\raggedright\begin{thebibliography}{100}

\bibitem{Flores_1994}
R.~A. Flores and J.~R. Primack, \emph{{Observational and theoretical
  constraints on singular dark matter halos}},
  \href{https://doi.org/10.1086/187350}{\emph{The Astrophysical Journal
  Letters} {\bfseries 427} (1994) L1}
  [\href{https://arxiv.org/abs/astro-ph/9402004}{{\ttfamily
  astro-ph/9402004}}].

\bibitem{Oh_2011}
S.-H. {Oh}, C.~{Brook}, F.~{Governato}, E.~{Brinks}, L.~{Mayer}, W.~J.~G. {de
  Blok}, A.~{Brooks} and F.~{Walter}, \emph{{The Central Slope of Dark Matter
  Cores in Dwarf Galaxies: Simulations versus {THINGS}}},
  \href{https://doi.org/10.1088/0004-6256/142/1/24}{\emph{The Astronomical
  Journal} {\bfseries 142} (2011) 24}
  [\href{https://arxiv.org/abs/1011.2777}{{\ttfamily 1011.2777}}].

\bibitem{walker2011method}
M.~G. {Walker} and J.~{Pe{\~n}arrubia}, \emph{{A Method for Measuring (Slopes
  of) the Mass Profiles of Dwarf Spheroidal Galaxies}},
  \href{https://doi.org/10.1088/0004-637X/742/1/20}{\emph{The Astrophysical
  Journal Letters} {\bfseries 742} (2011) 20}
  [\href{https://arxiv.org/abs/1108.2404}{{\ttfamily 1108.2404}}].

\bibitem{2011MNRAS.415L..40B}
M.~Boylan-Kolchin, J.~S. {Bullock} and M.~Kaplinghat, \emph{{Too big to fail?
  The puzzling darkness of massive Milky Way subhaloes}},
  \href{https://doi.org/10.1111/j.1745-3933.2011.01074.x}{\emph{Mon. Not. Roy.
  Astron. Soc.} {\bfseries 415} (2011) L40}
  [\href{https://arxiv.org/abs/1103.0007}{{\ttfamily 1103.0007}}].

\bibitem{DiValentino:2021izs}
E.~Di~Valentino, O.~Mena, S.~Pan, L.~Visinelli, W.~Yang, A.~Melchiorri, D.~F.
  Mota, A.~G. Riess and J.~Silk, \emph{{In the realm of the Hubble
  tension\textemdash{}a review of solutions}},
  \href{https://doi.org/10.1088/1361-6382/ac086d}{\emph{Class. Quant. Grav.}
  {\bfseries 38} (2021) 153001}
  [\href{https://arxiv.org/abs/2103.01183}{{\ttfamily 2103.01183}}].

\bibitem{Abdalla:2022yfr}
E.~Abdalla et~al., \emph{{Cosmology intertwined: A review of the particle
  physics, astrophysics, and cosmology associated with the cosmological
  tensions and anomalies}},
  \href{https://doi.org/10.1016/j.jheap.2022.04.002}{\emph{JHEAp} {\bfseries
  34} (2022) 49} [\href{https://arxiv.org/abs/2203.06142}{{\ttfamily
  2203.06142}}].

\bibitem{Blumenthal:1984bp}
G.~R. {Blumenthal}, S.~M. {Faber}, J.~R. {Primack} and M.~J. {Rees},
  \emph{{Formation of galaxies and large-scale structure with cold dark
  matter.}}, \href{https://doi.org/10.1038/311517a0}{\emph{Nature} {\bfseries
  311} (1984) 517}.

\bibitem{Bode:2000gq}
P.~Bode, J.~P. Ostriker and N.~Turok, \emph{{Halo formation in warm dark matter
  models}}, \href{https://doi.org/10.1086/321541}{\emph{The Astrophysical
  Journal} {\bfseries 556} (2001) 93}
  [\href{https://arxiv.org/abs/astro-ph/0010389}{{\ttfamily
  astro-ph/0010389}}].

\bibitem{Spergel:1999mh}
D.~N. Spergel and P.~J. Steinhardt, \emph{{Observational evidence for
  selfinteracting cold dark matter}},
  \href{https://doi.org/10.1103/PhysRevLett.84.3760}{\emph{Phys. Rev. Lett.}
  {\bfseries 84} (2000) 3760}
  [\href{https://arxiv.org/abs/astro-ph/9909386}{{\ttfamily
  astro-ph/9909386}}].

\bibitem{Vogelsberger:2014pda}
M.~Vogelsberger, J.~Zavala, C.~Simpson and A.~Jenkins, \emph{{Dwarf galaxies in
  CDM and SIDM with baryons: observational probes of the nature of dark
  matter}}, \href{https://doi.org/10.1093/mnras/stu1713}{\emph{Mon. Not. Roy.
  Astron. Soc.} {\bfseries 444} (2014) 3684}
  [\href{https://arxiv.org/abs/1405.5216}{{\ttfamily 1405.5216}}].

\bibitem{Tulin:2017ara}
S.~Tulin and H.-B. Yu, \emph{{Dark Matter Self-interactions and Small Scale
  Structure}}, \href{https://doi.org/10.1016/j.physrep.2017.11.004}{\emph{Phys.
  Rept.} {\bfseries 730} (2018) 1}
  [\href{https://arxiv.org/abs/1705.02358}{{\ttfamily 1705.02358}}].

\bibitem{Wang:2014ina}
M.-Y. Wang, A.~H.~G. Peter, L.~E. Strigari, A.~R. Zentner, B.~Arant,
  S.~Garrison-Kimmel and M.~Rocha, \emph{{Cosmological simulations of decaying
  dark matter: implications for small-scale structure of dark matter haloes}},
  \href{https://doi.org/10.1093/mnras/stu1747}{\emph{Mon. Not. Roy. Astron.
  Soc.} {\bfseries 445} (2014) 614}
  [\href{https://arxiv.org/abs/1406.0527}{{\ttfamily 1406.0527}}].

\bibitem{Hui:2016ltb}
L.~Hui, J.~P. Ostriker, S.~Tremaine and E.~Witten, \emph{{Ultralight scalars as
  cosmological dark matter}},
  \href{https://doi.org/10.1103/PhysRevD.95.043541}{\emph{Phys. Rev. D}
  {\bfseries 95} (2017) 043541}
  [\href{https://arxiv.org/abs/1610.08297}{{\ttfamily 1610.08297}}].

\bibitem{Du:2016zcv}
X.~Du, C.~Behrens and J.~C. Niemeyer, \emph{{Substructure of fuzzy dark matter
  haloes}}, \href{https://doi.org/10.1093/mnras/stw2724}{\emph{Mon. Not. Roy.
  Astron. Soc.} {\bfseries 465} (2017) 941}
  [\href{https://arxiv.org/abs/1608.02575}{{\ttfamily 1608.02575}}].

\bibitem{Maccio:2012qf}
A.~V. Maccio, S.~Paduroiu, D.~Anderhalden, A.~Schneider and B.~Moore,
  \emph{{Cores in warm dark matter haloes: a Catch 22 problem}},
  \href{https://doi.org/10.1111/j.1365-2966.2012.21284.x}{\emph{Mon. Not. Roy.
  Astron. Soc.} {\bfseries 424} (2012) 1105}
  [\href{https://arxiv.org/abs/1202.1282}{{\ttfamily 1202.1282}}].

\bibitem{Menci:2016eui}
N.~Menci, A.~Grazian, M.~Castellano and N.~G. Sanchez, \emph{{A Stringent Limit
  on the Warm Dark Matter Particle Masses from the Abundance of z=6 Galaxies in
  the Hubble Frontier Fields}},
  \href{https://doi.org/10.3847/2041-8205/825/1/L1}{\emph{The Astrophysical
  Journal Letters} {\bfseries 825} (2016) L1}
  [\href{https://arxiv.org/abs/1606.02530}{{\ttfamily 1606.02530}}].

\bibitem{Irsic:2017ixq}
V.~Ir\v{s}i\v{c} et~al., \emph{{New Constraints on the free-streaming of warm
  dark matter from intermediate and small scale Lyman-$\alpha$ forest data}},
  \href{https://doi.org/10.1103/PhysRevD.96.023522}{\emph{Phys. Rev. D}
  {\bfseries 96} (2017) 023522}
  [\href{https://arxiv.org/abs/1702.01764}{{\ttfamily 1702.01764}}].

\bibitem{Bowman:2018yin}
J.~D. Bowman, A.~E.~E. Rogers, R.~A. Monsalve, T.~J. Mozdzen and N.~Mahesh,
  \emph{An absorption profile centred at 78 megahertz in the sky-averaged
  spectrum}, \href{https://doi.org/10.1038/nature25792}{\emph{Nature}
  {\bfseries 555} (2018) 67}
  [\href{https://arxiv.org/abs/1810.05912}{{\ttfamily 1810.05912}}].

\bibitem{Barkana:2018lgd}
R.~Barkana, \emph{{Possible interaction between baryons and dark-matter
  particles revealed by the first stars}},
  \href{https://doi.org/10.1038/nature25791}{\emph{Nature} {\bfseries 555}
  (2018) 71} [\href{https://arxiv.org/abs/1803.06698}{{\ttfamily 1803.06698}}].

\bibitem{FURLANETTO2006181}
S.~R. {Furlanetto}, S.~P. {Oh} and F.~H. {Briggs}, \emph{{Cosmology at low
  frequencies: The 21 cm transition and the high-redshift Universe}},
  \href{https://doi.org/10.1016/j.physrep.2006.08.002}{\emph{Physics Reports}
  {\bfseries 433} (2006) 181}
  [\href{https://arxiv.org/abs/astro-ph/0608032}{{\ttfamily
  astro-ph/0608032}}].

\bibitem{2012RPPh...75h6901P}
J.~R. {Pritchard} and A.~{Loeb}, \emph{{21 cm cosmology in the 21st century}},
  \href{https://doi.org/10.1088/0034-4885/75/8/086901}{\emph{Reports on
  Progress in Physics} {\bfseries 75} (2012) 086901}
  [\href{https://arxiv.org/abs/1109.6012}{{\ttfamily 1109.6012}}].

\bibitem{fixsen2011arcade}
D.~J. {Fixsen}, A.~{Kogut}, S.~{Levin}, M.~{Limon}, P.~{Lubin}, P.~{Mirel},
  M.~{Seiffert}, J.~{Singal}, E.~{Wollack}, T.~{Villela} and C.~A. {Wuensche},
  \emph{{ARCADE 2 Measurement of the Absolute Sky Brightness at 3-90 GHz}},
  \href{https://doi.org/10.1088/0004-637X/734/1/5}{\emph{The Astrophysical
  Journal} {\bfseries 734} (2011) 5}
  [\href{https://arxiv.org/abs/0901.0555}{{\ttfamily 0901.0555}}].

\bibitem{2012JAI.....150004T}
G.~B. {Taylor}, S.~W. {Ellingson} et~al., \emph{{First Light for the First
  Station of the Long Wavelength Array}},
  \href{https://doi.org/10.1142/S2251171712500043}{\emph{Journal of
  Astronomical Instrumentation} {\bfseries 1} (2012) 1250004}
  [\href{https://arxiv.org/abs/1206.6733}{{\ttfamily 1206.6733}}].

\bibitem{Subrahmanyan:2013eqa}
R.~Subrahmanyan and R.~Cowsik, \emph{{Is there an Unaccounted for Excess in the
  Extragalactic Cosmic Radio Background?}},
  \href{https://doi.org/10.1088/0004-637X/776/1/42}{\emph{Astrophys. J.}
  {\bfseries 776} (2013) 42} [\href{https://arxiv.org/abs/1305.7060}{{\ttfamily
  1305.7060}}].

\bibitem{Dowell:2018mdb}
J.~Dowell and G.~B. Taylor, \emph{{The Radio Background Below 100 MHz}},
  \href{https://doi.org/10.3847/2041-8213/aabf86}{\emph{The Astrophysical
  Journal Letters} {\bfseries 858} (2018) L9}
  [\href{https://arxiv.org/abs/1804.08581}{{\ttfamily 1804.08581}}].

\bibitem{Fialkov:2019vnb}
A.~Fialkov and R.~Barkana, \emph{{Signature of Excess Radio Background in the
  21-cm Global Signal and Power Spectrum}},
  \href{https://doi.org/10.1093/mnras/stz873}{\emph{Mon. Not. Roy. Astron.
  Soc.} {\bfseries 486} (2019) 1763}
  [\href{https://arxiv.org/abs/1902.02438}{{\ttfamily 1902.02438}}].

\bibitem{Reis:2020arr}
I.~Reis, A.~Fialkov and R.~Barkana, \emph{{High-redshift radio galaxies: a
  potential new source of 21-cm fluctuations}},
  \href{https://doi.org/10.1093/mnras/staa3091}{\emph{Mon. Not. Roy. Astron.
  Soc.} {\bfseries 499} (2020) 5993}
  [\href{https://arxiv.org/abs/2008.04315}{{\ttfamily 2008.04315}}].

\bibitem{Feng:2018rje}
C.~Feng and G.~Holder, \emph{{Enhanced global signal of neutral hydrogen due to
  excess radiation at cosmic dawn}},
  \href{https://doi.org/10.3847/2041-8213/aac0fe}{\emph{Astrophys. J. Lett.}
  {\bfseries 858} (2018) L17}
  [\href{https://arxiv.org/abs/1802.07432}{{\ttfamily 1802.07432}}].

\bibitem{Sikder:2023ysk}
S.~Sikder, R.~Barkana, A.~Fialkov and I.~Reis, \emph{{Strong 21-cm fluctuations
  and anisotropy due to the line-of-sight effect of radio galaxies at cosmic
  dawn}},  \href{https://arxiv.org/abs/2301.04585}{{\ttfamily 2301.04585}}.

\bibitem{Fialkov:2018xre}
A.~Fialkov, R.~Barkana and A.~Cohen, \emph{{Constraining Baryon--Dark Matter
  Scattering with the Cosmic Dawn 21-cm Signal}},
  \href{https://doi.org/10.1103/PhysRevLett.121.011101}{\emph{Phys. Rev. Lett.}
  {\bfseries 121} (2018) 011101}
  [\href{https://arxiv.org/abs/1802.10577}{{\ttfamily 1802.10577}}].

\bibitem{Munoz:2018pzp}
J.~B. Mu\~noz and A.~Loeb, \emph{{A small amount of mini-charged dark matter
  could cool the baryons in the early Universe}},
  \href{https://doi.org/10.1038/s41586-018-0151-x}{\emph{Nature} {\bfseries
  557} (2018) 684} [\href{https://arxiv.org/abs/1802.10094}{{\ttfamily
  1802.10094}}].

\bibitem{Munoz:2018jwq}
J.~B. Mu\~noz, C.~Dvorkin and A.~Loeb, \emph{{21-cm Fluctuations from Charged
  Dark Matter}},
  \href{https://doi.org/10.1103/PhysRevLett.121.121301}{\emph{Phys. Rev. Lett.}
  {\bfseries 121} (2018) 121301}
  [\href{https://arxiv.org/abs/1804.01092}{{\ttfamily 1804.01092}}].

\bibitem{Mukhopadhyay:2021shy}
U.~{Mukhopadhyay}, D.~{Majumdar} and K.~K. {Datta}, \emph{{Probing interacting
  dark energy and scattering of baryons with dark matter in light of the EDGES
  21-cm signal}},
  \href{https://doi.org/10.1103/PhysRevD.103.063510}{\emph{Phys. Rev. D}
  {\bfseries 103} (2021) 063510}
  [\href{https://arxiv.org/abs/2008.09972}{{\ttfamily 2008.09972}}].

\bibitem{Aboubrahim:2021ohe}
A.~Aboubrahim, P.~Nath and Z.-Y. Wang, \emph{{A cosmologically consistent
  millicharged dark matter solution to the EDGES anomaly of possible string
  theory origin}}, \href{https://doi.org/10.1007/JHEP12(2021)148}{\emph{JHEP}
  {\bfseries 12} (2021) 148}
  [\href{https://arxiv.org/abs/2108.05819}{{\ttfamily 2108.05819}}].

\bibitem{Barkana:2022hko}
R.~Barkana, A.~Fialkov, H.~Liu and N.~J. Outmezguine, \emph{{Anticipating a
  New-Physics Signal in Upcoming 21-cm Power Spectrum Observations}},
  \href{https://arxiv.org/abs/2212.08082}{{\ttfamily 2212.08082}}.

\bibitem{silveira1985scalar}
V.~Silveira and A.~Zee, \emph{{Scalar Phantoms}},
  \href{https://doi.org/10.1016/0370-2693(85)90624-0}{\emph{Phys. Lett. B}
  {\bfseries 161} (1985) 136}.

\bibitem{mcdonald1994gauge}
J.~{McDonald}, \emph{{Gauge singlet scalars as cold dark matter}},
  \href{https://doi.org/10.1103/PhysRevD.50.3637}{\emph{Phys. Rev. D}
  {\bfseries 50} (1994) 3637}
  [\href{https://arxiv.org/abs/hep-ph/0702143}{{\ttfamily hep-ph/0702143}}].

\bibitem{burgess2001minimal}
C.~P. {Burgess}, M.~{Pospelov} and T.~{ter Veldhuis}, \emph{{The Minimal Model
  of nonbaryonic dark matter: a singlet scalar}},
  \href{https://doi.org/10.1016/S0550-3213(01)00513-2}{\emph{Nuclear Physics B}
  {\bfseries 619} (2001) 709}
  [\href{https://arxiv.org/abs/hep-ph/0011335}{{\ttfamily hep-ph/0011335}}].

\bibitem{cline2013update}
J.~M. {Cline}, P.~{Scott}, K.~{Kainulainen} and C.~{Weniger}, \emph{{Update on
  scalar singlet dark matter}},
  \href{https://doi.org/10.1103/PhysRevD.88.055025}{\emph{Phys. Rev. D}
  {\bfseries 88} (2013) 055025}
  [\href{https://arxiv.org/abs/1306.4710}{{\ttfamily 1306.4710}}].

\bibitem{feng2015closing}
L.~{Feng}, S.~{Profumo} and L.~{Ubaldi}, \emph{{Closing in on singlet scalar
  dark matter: LUX, invisible Higgs decays and gamma-ray lines}},
  \href{https://doi.org/10.1007/JHEP03(2015)045}{\emph{Journal of High Energy
  Physics} {\bfseries 2015} (2015) 45}
  [\href{https://arxiv.org/abs/1412.1105}{{\ttfamily 1412.1105}}].

\bibitem{VanDong:2022rox}
P.~Van~Dong, \emph{{Physics implication from a Z3 symmetry of matter}},
  \href{https://doi.org/10.1103/PhysRevD.107.055026}{\emph{Phys. Rev. D}
  {\bfseries 107} (2023) 055026}
  [\href{https://arxiv.org/abs/2205.04253}{{\ttfamily 2205.04253}}].

\bibitem{koerich2015dark}
L.~V. Koerich, \emph{Dark matter in a $\mathbb{Z}_3$-symmetry extension of the
  standard model}, .

\bibitem{Belanger:2012zr}
G.~Belanger, K.~Kannike, A.~Pukhov and M.~Raidal, \emph{{$Z_3$ Scalar Singlet
  Dark Matter}},
  \href{https://doi.org/10.1088/1475-7516/2013/01/022}{\emph{JCAP} {\bfseries
  01} (2013) 022} [\href{https://arxiv.org/abs/1211.1014}{{\ttfamily
  1211.1014}}].

\bibitem{Bernal:2015lbl}
N.~{Bernal}, C.~{Garcia-Cely} and R.~{Rosenfeld}, \emph{{Z$_{3}$ WIMP and SIMP
  Dark Matter from a Global U(1) Breaking}},
  \href{https://doi.org/10.1016/j.nuclphysbps.2015.11.001}{\emph{Nuclear and
  Particle Physics Proceedings} {\bfseries 267-269} (2015) 353}.

\bibitem{DiazSaez:2022nhp}
B.~D\'\i{}az~S\'aez and K.~Ghorbani, \emph{{Z $_{3}$ scalar dark matter with
  strong positron fluxes}},
  \href{https://doi.org/10.1088/1475-7516/2023/02/002}{\emph{JCAP} {\bfseries
  02} (2023) 002} [\href{https://arxiv.org/abs/2203.09282}{{\ttfamily
  2203.09282}}].

\bibitem{DEramo:2010keq}
F.~D'Eramo and J.~Thaler, \emph{{Semi-annihilation of Dark Matter}},
  \href{https://doi.org/10.1007/JHEP06(2010)109}{\emph{JHEP} {\bfseries 06}
  (2010) 109} [\href{https://arxiv.org/abs/1003.5912}{{\ttfamily 1003.5912}}].

\bibitem{DEramo:2011rlb}
F.~D'Eramo, \emph{{Semi-annihilation of Dark Matter}},
  \href{https://doi.org/10.22323/1.110.0083}{\emph{PoS} {\bfseries IDM2010}
  (2011) 083} [\href{https://arxiv.org/abs/1101.5413}{{\ttfamily 1101.5413}}].

\bibitem{hambye2009hidden}
T.~{Hambye} and M.~H.~G. {Tytgat}, \emph{{Confined hidden vector dark matter}},
  \href{https://doi.org/10.1016/j.physletb.2009.11.050}{\emph{Physics Letters
  B} {\bfseries 683} (2010) 39}
  [\href{https://arxiv.org/abs/0907.1007}{{\ttfamily 0907.1007}}].

\bibitem{Belanger:2012vp}
G.~Belanger, K.~Kannike, A.~Pukhov and M.~Raidal, \emph{{Impact of
  semi-annihilations on dark matter phenomenology - an example of $Z_N$
  symmetric scalar dark matter}},
  \href{https://doi.org/10.1088/1475-7516/2012/04/010}{\emph{JCAP} {\bfseries
  04} (2012) 010} [\href{https://arxiv.org/abs/1202.2962}{{\ttfamily
  1202.2962}}].

\bibitem{Belanger:2014bga}
G.~B\'elanger, K.~Kannike, A.~Pukhov and M.~Raidal, \emph{{Minimal
  semi-annihilating $\mathbb{Z}_N$ scalar dark matter}},
  \href{https://doi.org/10.1088/1475-7516/2014/06/021}{\emph{JCAP} {\bfseries
  06} (2014) 021} [\href{https://arxiv.org/abs/1403.4960}{{\ttfamily
  1403.4960}}].

\bibitem{Bandyopadhyay:2022tsf}
P.~Bandyopadhyay, D.~Choudhury and D.~Sachdeva, \emph{{Semiannihilation of
  fermionic dark matter}},
  \href{https://doi.org/10.1103/PhysRevD.107.015020}{\emph{Phys. Rev. D}
  {\bfseries 107} (2023) 015020}
  [\href{https://arxiv.org/abs/2206.05811}{{\ttfamily 2206.05811}}].

\bibitem{Ghosh:2020lma}
A.~Ghosh, D.~Ghosh and S.~Mukhopadhyay, \emph{{Asymmetric dark matter from
  semi-annihilation}},
  \href{https://doi.org/10.1007/JHEP08(2020)149}{\emph{JHEP} {\bfseries 08}
  (2020) 149} [\href{https://arxiv.org/abs/2004.07705}{{\ttfamily
  2004.07705}}].

\bibitem{Hochberg:2014dra}
Y.~Hochberg, E.~Kuflik, T.~Volansky and J.~G. Wacker, \emph{{Mechanism for
  Thermal Relic Dark Matter of Strongly Interacting Massive Particles}},
  \href{https://doi.org/10.1103/PhysRevLett.113.171301}{\emph{Phys. Rev. Lett.}
  {\bfseries 113} (2014) 171301}
  [\href{https://arxiv.org/abs/1402.5143}{{\ttfamily 1402.5143}}].

\bibitem{Hochberg:2014kqa}
Y.~Hochberg, E.~Kuflik, H.~Murayama, T.~Volansky and J.~G. Wacker, \emph{{Model
  for Thermal Relic Dark Matter of Strongly Interacting Massive Particles}},
  \href{https://doi.org/10.1103/PhysRevLett.115.021301}{\emph{Phys. Rev. Lett.}
  {\bfseries 115} (2015) 021301}
  [\href{https://arxiv.org/abs/1411.3727}{{\ttfamily 1411.3727}}].

\bibitem{Teplitz:2000zd}
V.~L. Teplitz, R.~N. Mohapatra, F.~I. Olness and R.~Stroynowski, \emph{{SIMP
  (Strongly Interacting Massive Particle) search}},  in \emph{{4th
  International Symposium on Sources and Detection of Dark Matter in the
  Universe (DM 2000)}}, pp.~256--262, 2, 2000,
  \href{https://arxiv.org/abs/hep-ph/0005111}{{\ttfamily hep-ph/0005111}}.

\bibitem{Mohapatra:1999gg}
R.~N. Mohapatra, F.~I. Olness, R.~Stroynowski and V.~L. Teplitz,
  \emph{{Searching for strongly interacting massive particles (SIMPs)}},
  \href{https://doi.org/10.1103/PhysRevD.60.115013}{\emph{Phys. Rev. D}
  {\bfseries 60} (1999) 115013}
  [\href{https://arxiv.org/abs/hep-ph/9906421}{{\ttfamily hep-ph/9906421}}].

\bibitem{Smirnov:2020zwf}
J.~Smirnov and J.~F. Beacom, \emph{{New Freezeout Mechanism for Strongly
  Interacting Dark Matter}},
  \href{https://doi.org/10.1103/PhysRevLett.125.131301}{\emph{Phys. Rev. Lett.}
  {\bfseries 125} (2020) 131301}
  [\href{https://arxiv.org/abs/2002.04038}{{\ttfamily 2002.04038}}].

\bibitem{Parikh:2023qtk}
A.~Parikh, J.~Smirnov, W.~L. Xu and B.~Zhou, \emph{{Scalar Co-SIMP Dark Matter:
  Models and Sensitivities}},
  \href{https://arxiv.org/abs/2302.00008}{{\ttfamily 2302.00008}}.

\bibitem{Marsicano:2018vin}
L.~Marsicano, M.~Battaglieri, A.~Celentano, R.~De~Vita and Y.-M. Zhong,
  \emph{{Probing Leptophilic Dark Sectors at Electron Beam-Dump Facilities}},
  \href{https://doi.org/10.1103/PhysRevD.98.115022}{\emph{Phys. Rev. D}
  {\bfseries 98} (2018) 115022}
  [\href{https://arxiv.org/abs/1812.03829}{{\ttfamily 1812.03829}}].

\bibitem{Batell:2014mga}
B.~Batell, R.~Essig and Z.~Surujon, \emph{{Strong Constraints on Sub-GeV Dark
  Sectors from SLAC Beam Dump E137}},
  \href{https://doi.org/10.1103/PhysRevLett.113.171802}{\emph{Phys. Rev. Lett.}
  {\bfseries 113} (2014) 171802}
  [\href{https://arxiv.org/abs/1406.2698}{{\ttfamily 1406.2698}}].

\bibitem{Muong-2:2021ojo}
{\scshape Muon g-2} Collaboration, B.~Abi et~al., \emph{{Measurement of the
  Positive Muon Anomalous Magnetic Moment to 0.46 ppm}},
  \href{https://doi.org/10.1103/PhysRevLett.126.141801}{\emph{Phys. Rev. Lett.}
  {\bfseries 126} (2021) 141801}
  [\href{https://arxiv.org/abs/2104.03281}{{\ttfamily 2104.03281}}].

\bibitem{Seager:1999bc}
S.~Seager, D.~D. Sasselov and D.~Scott, \emph{{A new calculation of the
  recombination epoch}}, \href{https://doi.org/10.1086/312250}{\emph{The
  Astrophysical Journal Letters} {\bfseries 523} (1999) L1}
  [\href{https://arxiv.org/abs/astro-ph/9909275}{{\ttfamily
  astro-ph/9909275}}].

\bibitem{Singh:2021mxo}
S.~Singh, J.~Nambissan~T., R.~Subrahmanyan, N.~Udaya~Shankar, B.~S. Girish,
  A.~Raghunathan, R.~Somashekar, K.~S. Srivani and M.~Sathyanarayana~Rao,
  \emph{{On the detection of a cosmic dawn signal in the radio background}},
  \href{https://doi.org/10.1038/s41550-022-01610-5}{\emph{Nature Astron.}
  {\bfseries 6} (2022) 607} [\href{https://arxiv.org/abs/2112.06778}{{\ttfamily
  2112.06778}}].

\bibitem{Bevins:2022ajf}
H.~T.~J. Bevins, A.~Fialkov, E.~d.~L. Acedo, W.~J. Handley, S.~Singh,
  R.~Subrahmanyan and R.~Barkana, \emph{{Astrophysical constraints from the
  SARAS 3 non-detection of the cosmic dawn sky-averaged 21-cm signal}},
  \href{https://doi.org/10.1038/s41550-022-01825-6}{\emph{Nature Astron.}
  {\bfseries 6} (2022) 1473}
  [\href{https://arxiv.org/abs/2212.00464}{{\ttfamily 2212.00464}}].

\bibitem{2015aska.confE.174B}
R.~{Braun}, T.~{Bourke}, J.~A. {Green}, E.~{Keane} and J.~{Wagg},
  \emph{{Advancing Astrophysics with the Square Kilometre Array}}, .

\bibitem{bale2023lusee}
S.~D. Bale, N.~Bassett, J.~O. Burns, J.~D. Jones, K.~Goetz, C.~Hellum-Bye,
  S.~Hermann, J.~Hibbard, M.~Maksimovic, R.~McLean, R.~Monsalve, P.~O'Connor,
  A.~Parsons, M.~Pulupa, R.~Pund, D.~Rapetti, K.~M. Rotermund, B.~Saliwanchik,
  A.~Slosar, D.~Sundkvist and A.~Suzuki, \emph{Lusee 'night': The lunar surface
  electromagnetics experiment},  2023.

\bibitem{DeBoer:2016tnn}
D.~R. DeBoer et~al., \emph{{Hydrogen Epoch of Reionization Array (HERA)}},
  \href{https://doi.org/10.1088/1538-3873/129/974/045001}{\emph{Publ. Astron.
  Soc. Pac.} {\bfseries 129} (2017) 045001}
  [\href{https://arxiv.org/abs/1606.07473}{{\ttfamily 1606.07473}}].

\bibitem{Tingay:2012ps}
S.~J. Tingay et~al., \emph{{The Murchison Widefield Array: the Square Kilometre
  Array Precursor at low radio frequencies}},
  \href{https://doi.org/10.1017/pasa.2012.007}{\emph{Publ. Astron. Soc.
  Austral.} {\bfseries 30} (2013) 7}
  [\href{https://arxiv.org/abs/1206.6945}{{\ttfamily 1206.6945}}].

\bibitem{8471648}
P.~Zarka, A.~Coffre, L.~Denis, C.~Dumez-Viou, J.~Girard, J.-M. Grießmeier,
  A.~Loh and M.~Tagger, \emph{The low-frequency radiotelescope nenufar},  in
  \emph{2018 2nd URSI Atlantic Radio Science Meeting (AT-RASC)}, pp.~1--1,
  2018, \href{https://doi.org/10.23919/URSI-AT-RASC.2018.8471648}{DOI}.

\bibitem{deLeraAcedo:2022kiu}
E.~de~Lera~Acedo et~al., \emph{{The REACH radiometer for detecting the 21-cm
  hydrogen signal from redshift z\,\ensuremath{\approx}\,7.5\textendash{}28}},
  \href{https://doi.org/10.1038/s41550-022-01817-6}{\emph{Nature Astron.}
  {\bfseries 6} (2022) 998} [\href{https://arxiv.org/abs/2210.07409}{{\ttfamily
  2210.07409}}].

\bibitem{MIST}
\emph{``mist project homepage''}, . \url{http://www.physics.mcgill.ca/mist/}.

\bibitem{Chen:2019xvd}
X.~Chen et~al., \emph{{Discovering the Sky at the Longest Wavelengths with
  Small Satellite Constellations}},  in \emph{{ISSI-BJ Forum}: {Discover the
  Sky by Longest Wavelength with Small Satellite Constellation}}, 7, 2019,
  \href{https://arxiv.org/abs/1907.10853}{{\ttfamily 1907.10853}}.

\bibitem{pratush}
\emph{``pratush project homepage''}, .
  \url{https://wwws.rri.res.in/DISTORTION/pratush.html}.

\bibitem{2011JCAP...07..034B}
D.~{Blas}, J.~{Lesgourgues} and T.~{Tram}, \emph{{The Cosmic Linear Anisotropy
  Solving System (CLASS). Part II: Approximation schemes}},
  \href{https://doi.org/10.1088/1475-7516/2011/07/034}{\emph{The Astrophysical
  Journal} {\bfseries 2011} (2011) 034}
  [\href{https://arxiv.org/abs/1104.2933}{{\ttfamily 1104.2933}}].

\bibitem{Brinckmann:2018cvx}
T.~Brinckmann and J.~Lesgourgues, \emph{{MontePython 3: boosted MCMC sampler
  and other features}},
  \href{https://doi.org/10.1016/j.dark.2018.100260}{\emph{Phys. Dark Univ.}
  {\bfseries 24} (2019) 100260}
  [\href{https://arxiv.org/abs/1804.07261}{{\ttfamily 1804.07261}}].

\bibitem{Planck:2018vyg}
{\scshape Planck} Collaboration, N.~Aghanim et~al., \emph{{Planck 2018 results.
  VI. Cosmological parameters}},
  \href{https://doi.org/10.1051/0004-6361/201833910}{\emph{Astron. Astrophys.}
  {\bfseries 641} (2020) A6}
  [\href{https://arxiv.org/abs/1807.06209}{{\ttfamily 1807.06209}}].

\bibitem{1958PIRE...46..240F}
G.~B. {Field}, \emph{{Excitation of the Hydrogen 21-CM Line}},
  \href{https://doi.org/10.1109/JRPROC.1958.286741}{\emph{Proceedings of the
  IRE} {\bfseries 46} (1958) 240}.

\bibitem{1959ApJ...129..551F}
G.~B. {Field}, \emph{{The Time Relaxation of a Resonance-Line Profile.}},
  \href{https://doi.org/10.1086/146654}{\emph{The Astrophysical Journal}
  {\bfseries 129} (1959) 551}.

\bibitem{Kuhlen_2006}
M.~{Kuhlen}, P.~{Madau} and R.~{Montgomery}, \emph{{The Spin Temperature and 21
  cm Brightness of the Intergalactic Medium in the Pre-Reionization era}},
  \href{https://doi.org/10.1086/500548}{\emph{The Astrophysical Journal
  Letters} {\bfseries 637} (2006) L1}
  [\href{https://arxiv.org/abs/astro-ph/0510814}{{\ttfamily
  astro-ph/0510814}}].

\bibitem{PhysRevD.74.103502}
S.~R. {Furlanetto}, S.~P. {Oh} and E.~{Pierpaoli}, \emph{{Effects of dark
  matter decay and annihilation on the high-redshift 21cm background}},
  \href{https://doi.org/10.1103/PhysRevD.74.103502}{\emph{Phys. Rev. D}
  {\bfseries 74} (2006) 103502}
  [\href{https://arxiv.org/abs/astro-ph/0608385}{{\ttfamily
  astro-ph/0608385}}].

\bibitem{Pritchard_2006}
J.~R. {Pritchard} and S.~R. {Furlanetto}, \emph{{Descending from on high:
  Lyman-series cascades and spin-kinetic temperature coupling in the 21-cm
  line}}, \href{https://doi.org/10.1111/j.1365-2966.2006.10028.x}{\emph{Mon.
  Not. Roy. Astron. Soc.} {\bfseries 367} (2006) 1057}
  [\href{https://arxiv.org/abs/astro-ph/0508381}{{\ttfamily
  astro-ph/0508381}}].

\bibitem{Barkana_2005}
R.~{Barkana} and A.~{Loeb}, \emph{{Detecting the Earliest Galaxies through Two
  New Sources of 21 Centimeter Fluctuations}},
  \href{https://doi.org/10.1086/429954}{\emph{The Astrophysical Journal}
  {\bfseries 626} (2005) 1}
  [\href{https://arxiv.org/abs/astro-ph/0410129}{{\ttfamily
  astro-ph/0410129}}].

\bibitem{Bharadwaj_2004}
S.~{Bharadwaj} and S.~S. {Ali}, \emph{{The cosmic microwave background
  radiation fluctuations from HI perturbations prior to reionization}},
  \href{https://doi.org/10.1111/j.1365-2966.2004.07907.x}{\emph{Mon. Not. Roy.
  Astron. Soc.} {\bfseries 352} (2004) 142}
  [\href{https://arxiv.org/abs/astro-ph/0401206}{{\ttfamily
  astro-ph/0401206}}].

\bibitem{Aver:2015iza}
E.~Aver, K.~A. Olive and E.~D. Skillman, \emph{{The effects of He I
  \ensuremath{\lambda}10830 on helium abundance determinations}},
  \href{https://doi.org/10.1088/1475-7516/2015/07/011}{\emph{JCAP} {\bfseries
  07} (2015) 011} [\href{https://arxiv.org/abs/1503.08146}{{\ttfamily
  1503.08146}}].

\bibitem{Chatterjee:2019jts}
A.~Chatterjee, P.~Dayal, T.~R. Choudhury and A.~Hutter, \emph{{Ruling out 3 keV
  warm dark matter using 21 cm EDGES data}},
  \href{https://doi.org/10.1093/mnras/stz1444}{\emph{Mon. Not. Roy. Astron.
  Soc.} {\bfseries 487} (2019) 3560}
  [\href{https://arxiv.org/abs/1902.09562}{{\ttfamily 1902.09562}}].

\bibitem{Schneider:2018xba}
A.~Schneider, \emph{{Constraining noncold dark matter models with the global
  21-cm signal}}, \href{https://doi.org/10.1103/PhysRevD.98.063021}{\emph{Phys.
  Rev. D} {\bfseries 98} (2018) 063021}
  [\href{https://arxiv.org/abs/1805.00021}{{\ttfamily 1805.00021}}].

\bibitem{Wise_2014}
J.~H. Wise, V.~G. Demchenko, M.~T. Halicek, M.~L. Norman, M.~J. Turk, T.~Abel
  and B.~D. Smith, \emph{{The birth of a galaxy \textendash{} III. Propelling
  reionization with the faintest galaxies}},
  \href{https://doi.org/10.1093/mnras/stu979}{\emph{Mon. Not. Roy. Astron.
  Soc.} {\bfseries 442} (2014) 2560}
  [\href{https://arxiv.org/abs/1403.6123}{{\ttfamily 1403.6123}}].

\bibitem{2003MNRAS.340..210G}
S.~C.~O. {Glover} and P.~W.~J.~L. {Brand}, \emph{{Radiative feedback from an
  early X-ray background}},
  \href{https://doi.org/10.1046/j.1365-8711.2003.06311.x}{\emph{Mon. Not. Roy.
  Astron. Soc.} {\bfseries 340} (2003) 210}
  [\href{https://arxiv.org/abs/astro-ph/0205308}{{\ttfamily
  astro-ph/0205308}}].

\bibitem{Chluba_2010}
J.~{Chluba}, \emph{{Could the cosmological recombination spectrum help us
  understand annihilating dark matter?}},
  \href{https://doi.org/10.1111/j.1365-2966.2009.15957.x}{\emph{Mon. Not. Roy.
  Astron. Soc.} {\bfseries 402} (2010) 1195}
  [\href{https://arxiv.org/abs/0910.3663}{{\ttfamily 0910.3663}}].

\bibitem{feng2018enhanced}
C.~{Feng} and G.~{Holder}, \emph{{Enhanced Global Signal of Neutral Hydrogen
  Due to Excess Radiation at Cosmic Dawn}},
  \href{https://doi.org/10.3847/2041-8213/aac0fe}{\emph{The Astrophysical
  Journal Letters} {\bfseries 858} (2018) L17}
  [\href{https://arxiv.org/abs/1802.07432}{{\ttfamily 1802.07432}}].

\bibitem{Tashiro:2014tsa}
H.~Tashiro, K.~Kadota and J.~Silk, \emph{{Effects of dark matter-baryon
  scattering on redshifted 21 cm signals}},
  \href{https://doi.org/10.1103/PhysRevD.90.083522}{\emph{Phys. Rev. D}
  {\bfseries 90} (2014) 083522}
  [\href{https://arxiv.org/abs/1408.2571}{{\ttfamily 1408.2571}}].

\bibitem{2021arXiv210401756N}
J.~{Nambissan T.}, R.~{Subrahmanyan}, R.~{Somashekar}, N.~{Udaya Shankar},
  S.~{Singh}, A.~{Raghunathan}, B.~S. {Girish}, K.~S. {Srivani} and
  M.~{Sathyanarayana Rao}, \emph{{SARAS 3 CD/EoR Radiometer: Design and
  Performance of the Receiver}},
  \href{https://doi.org/10.48550/arXiv.2104.01756}{\emph{arXiv e-prints} (2021)
  arXiv:2104.01756} [\href{https://arxiv.org/abs/2104.01756}{{\ttfamily
  2104.01756}}].

\bibitem{Girish_2020}
B.~S. Girish, K.~S. Srivani, R.~Subrahmanyan, N.~U. Shankar, S.~Singh, T.~J.
  Nambissan, M.~S. Rao, R.~Somashekar and A.~Raghunathan, \emph{{SARAS}
  {CD}/{EoR} radiometer: Design and performance of the digital correlation
  spectrometer}, \href{https://doi.org/10.1142/s2251171720500063}{\emph{Journal
  of Astronomical Instrumentation} {\bfseries 09} (2020) }.

\bibitem{Singh:2017gtp}
S.~Singh et~al., \emph{{First results on the Epoch of Reionization from First
  Light with SARAS 2}},
  \href{https://doi.org/10.3847/2041-8213/aa831b}{\emph{The Astrophysical
  Journal Letters} {\bfseries 845} (2017) L12}
  [\href{https://arxiv.org/abs/1703.06647}{{\ttfamily 1703.06647}}].

\bibitem{Patra_2013}
N.~Patra, R.~Subrahmanyan, A.~Raghunathan and N.~U. Shankar, \emph{{SARAS}: a
  precision system for measurement of the cosmic radio background and
  signatures from the epoch of reionization},
  \href{https://doi.org/10.1007/s10686-013-9336-3}{\emph{Experimental
  Astronomy} {\bfseries 36} (2013) 319}
  [\href{https://arxiv.org/abs/1211.3800}{{\ttfamily 1211.3800}}].

\bibitem{Mondal:2023xjx}
R.~{Mondal} and R.~{Barkana}, \emph{{Precision cosmology with the 21-cm signal
  from the dark ages}},
  \href{https://doi.org/10.48550/arXiv.2305.08593}{\emph{arXiv e-prints} (2023)
  arXiv:2305.08593} [\href{https://arxiv.org/abs/2305.08593}{{\ttfamily
  2305.08593}}].

\bibitem{Jana:2018gqk}
R.~Jana, B.~B. Nath and P.~L. Biermann, \emph{{Radio background and IGM heating
  due to Pop III supernova explosions}},
  \href{https://doi.org/10.1093/mnras/sty3426}{\emph{Mon. Not. Roy. Astron.
  Soc.} {\bfseries 483} (2019) 5329}
  [\href{https://arxiv.org/abs/1812.07404}{{\ttfamily 1812.07404}}].

\bibitem{Mebane:2020jwl}
R.~H. Mebane, J.~Mirocha and S.~R. Furlanetto, \emph{{The effects of population
  III radiation backgrounds on the cosmological 21-cm signal}},
  \href{https://doi.org/10.1093/mnras/staa280}{\emph{Mon. Not. Roy. Astron.
  Soc.} {\bfseries 493} (2020) 1217}
  [\href{https://arxiv.org/abs/1910.10171}{{\ttfamily 1910.10171}}].

\bibitem{Matsumura:2013aja}
T.~Matsumura et~al., \emph{{Mission design of LiteBIRD}},
  \href{https://doi.org/10.1007/s10909-013-0996-1}{\emph{J. Low Temp. Phys.}
  {\bfseries 176} (2014) 733}
  [\href{https://arxiv.org/abs/1311.2847}{{\ttfamily 1311.2847}}].

\bibitem{Suzuki:2018cuy}
A.~Suzuki et~al., \emph{{The LiteBIRD Satellite Mission - Sub-Kelvin
  Instrument}}, \href{https://doi.org/10.1007/s10909-018-1947-7}{\emph{J. Low
  Temp. Phys.} {\bfseries 193} (2018) 1048}
  [\href{https://arxiv.org/abs/1801.06987}{{\ttfamily 1801.06987}}].

\bibitem{CMB-S4:2016ple}
{\scshape CMB-S4} Collaboration, K.~N. Abazajian et~al., \emph{{CMB-S4 Science
  Book, First Edition}},  \href{https://arxiv.org/abs/1610.02743}{{\ttfamily
  1610.02743}}.

\bibitem{CMB-S4:2017uhf}
{\scshape CMB-S4} Collaboration, M.~H. Abitbol et~al., \emph{{CMB-S4 Technology
  Book, First Edition}},  \href{https://arxiv.org/abs/1706.02464}{{\ttfamily
  1706.02464}}.

\bibitem{EuclidTheoryWorkingGroup:2012gxx}
{\scshape Euclid Theory Working Group} Collaboration, L.~Amendola et~al.,
  \emph{{Cosmology and fundamental physics with the Euclid satellite}},
  \href{https://doi.org/10.12942/lrr-2013-6}{\emph{Living Rev. Rel.} {\bfseries
  16} (2013) 6} [\href{https://arxiv.org/abs/1206.1225}{{\ttfamily
  1206.1225}}].

\bibitem{DESI:2013agm}
{\scshape DESI} Collaboration, M.~Levi et~al., \emph{{The DESI Experiment, a
  whitepaper for Snowmass 2013}},
  \href{https://arxiv.org/abs/1308.0847}{{\ttfamily 1308.0847}}.

\bibitem{DESI:2016fyo}
{\scshape DESI} Collaboration, A.~Aghamousa et~al., \emph{{The DESI Experiment
  Part I: Science,Targeting, and Survey Design}},
  \href{https://arxiv.org/abs/1611.00036}{{\ttfamily 1611.00036}}.

\end{thebibliography}\endgroup

\end{document}